%% file: ms07.tex
\documentclass[usenatbib]{mn2e}

\usepackage{natbib}
\usepackage{fixltx2e}
\usepackage{amsmath}
\usepackage{times}
\usepackage[pdftex]{graphicx}
\usepackage{multirow}
\usepackage{float}

\input{defn}
\providecommand{\e}[1]{\ensuremath{\times 10^{#1}}}

\providecommand{\err}[2]{\ensuremath{^{+#1}_{-#2}}}

\begin{document}

\title[AGN Cycling in MS0735]{Cycling of the powerful AGN in MS 0735.6+7421 and the duty cycle of radio AGN in Clusters}
\author[ A.~N. Vantyghem et al.]  
    {\parbox[]{7.in}{A.~N. Vantyghem$^1$, B.~R. McNamara$^{1,2,3}$, H.~R. Russell$^{1}$, R.~A. Main$^1$, P.~E.~J. Nulsen$^3$, M.~W. Wise$^4$, H. Hoekstra$^5$, M. Gitti$^{6,7}$ \\
    \footnotesize 
    $^1$ Department of Physics and Astronomy, University of Waterloo, Waterloo, ON N2L 3G1, Canada \\ 
    $^2$ Perimeter Institute for Theoretical Physics, Waterloo, Canada \\
    $^3$ Harvard-Smithsonian Center for Astrophysics, 60 Garden Street, Cambridge, MA 02138, USA \\
    $^4$ ASTRON (Netherlands Institute for Radio Astronomy), PO Box 2, 7990 AA Dwingeloo, the Netherlands \\
    $^5$ Leiden Observatory, Leiden University, PO Box 9513, 2300 RA Leiden, the Netherlands \\
    $^6$ Dipartimento di Fisica e Astronomia - Universit\`a di Bologna, via Ranzani 1, I-40127 Bologna, Italy \\
    $^7$ INAF - Istituto di Radioastronomia, via Gobetti 101, I-40129 Bologna, Italy
  }
}
\maketitle

\begin{abstract}

We present an analysis of deep \textit{Chandra} X-ray observations of the galaxy cluster MS 0735.6+7421, which hosts the most energetic radio AGN known. Our analysis has revealed two cavities in its hot atmosphere with diameters of $200-240\kpc$. The total cavity enthalpy, mean age, and mean jet power are $9\e{61}\erg$, $1.6\e{8}\yr$, and $1.7\e{46}\ergps$, respectively. The cavities are surrounded by nearly continuous temperature and surface brightness discontinuities associated with an elliptical shock front of Mach number $1.26$ ($1.17-1.30$) and age of $1.1\e{8}\yr$. The shock has injected at least $4\e{61}\erg$ into the hot atmosphere at a rate of $1.1\e{46}\ergps$. A second pair of cavities and possibly a second shock front are located along the radio jets, indicating that the AGN power has declined by a factor of $30$ over the past $100\Myr$. The multiphase atmosphere surrounding the central galaxy is cooling at a rate of $40\Msunpyr$, but does not fuel star formation at an appreciable rate. In addition to heating, entrainment in the radio jet may be depleting the nucleus of fuel and preventing gas from condensing out of the intracluster medium. Finally, we examine the mean time intervals between AGN outbursts in systems with multiple generations of X-ray cavities. We find that, like MS0735, their AGN rejuvenate on a timescale that is approximately $1/3$ of their mean central cooling timescales, indicating that jet heating is outpacing cooling in these systems.

\end{abstract}


\section{Introduction}

The hot atmospheres at the centres of cool core clusters are often bright enough to radiate away their energy in 
X-rays in less than $10^9 \yr$. If uncompensated by heating, the atmospheres will cool and fuel star formation in 
central cluster galaxies at rates of tens to hundreds of solar masses per year (reviewed by \citealt{fabian94}). 
However, high resolution spectroscopy with the \textit{Chandra} and \textit{XMM-Newton} X-ray Observatories revealed 
only weak X-ray line-emission below $1\keV$ that is inconsistent with gas cooling out of the X-ray band at the 
expected rates, implying that cooling is compensated by heating \citep{peterson03}. The most likely heating mechanism 
is feedback from the central active galactic nucleus (AGN) (reviewed by \citealt{review07}). Outbursts from the 
central AGN inflate bubbles filled with radio emission that are visible as surface brightness depressions, or 
cavities, in X-ray imaging. These bubbles heat the ICM in their wake as they rise buoyantly through the cluster 
atmosphere \citep{churazov01}. A study of the Brightest 55 clusters showed that bubbles are present in at least 
$70\%$ of cool core clusters \citep{dunn06,birzan12}, though this fraction may actually exceed $95\%$ \citep{fabian12}.

Radio jets launched by central AGN also drive shock fronts into hot atmospheres that have been identified in a growing
number of clusters and groups (e.g. M87: \citealt{forman07}, Hydra A: \citealt{nulsenhydra}, A2052: \citealt{blanton11},
Hercules A: \citealt{nulsenherc}, NGC 5813: \citealt{randall11}, and others). Though their total energies can be large,
the shocks are usually weak with Mach numbers lying between $1.2$ and $1.7$. Heating from weak shocks is most effective
at small radii, and may be a critical element of a feedback cycle (reviewed by \citealt{review12}). Sound waves, such as
those detected in Perseus \citep{fabian06}, Centaurus \citep{sanders08}, and A2052 \citep{blanton11}, deposit energy on
large scales. Quantifying their contribution to heating is difficult as it depends on uncertain transport coefficients.

Continual AGN activity is required to suppress cooling over the ages of clusters. AGN must rejuvenate on timescales
shorter than the central cooling time in order to regulate or prevent star formation in central galaxies. Sequential 
AGN outbursts are indicated by the radio morphologies of several systems \citep{schoenmakers1, saripalli02}. Likewise,
deep \textit{Chandra} observations have revealed multiple generations of cavities from a number of nearby systems,
including Perseus \citep{fabian06}, M87 \citep{forman07}, Hydra A \citep{wise07}, HCG 62 \citep{rafferty13}, Abell 2199
\citep{nulsen13}, Abell 2052 \citep{blanton09, blanton11}, NGC 5813 \citep{randall11}, Abell 3581 \citep{canning13}, 
and NGC 5846 \citep{machacek11}.

The cool core cluster MS 0735.6+7421 (hereafter MS0735) hosts unusually large X-ray cavities in an otherwise relaxed
system \citep{gitti07}. Each cavity has a diamater of roughly $200\kpc$ and is filled with synchrotron emission from the
radio jet. A weak but powerful shock front encompasses the cavities \citep{nature05}. The total energy required to
inflate the bubbles and drive the shock front exceeds $10^{62}\erg$, making this the most powerful AGN outburst known.
The interaction between radio jet and the surrounding hot atmosphere is a key piece of information for the process of 
AGN feedback. Deep X-ray observations of the most powerful AGN outbursts are crucial in understanding this interaction.

MS0735 is an enigmatic object that challenges our understanding of how galaxies and supermassive black holes coevolve 
and how AGN are powered. With a mechanical power in excess of $10^{46}\ergps$, quasar-like power output requires 
$\sim5\e{8}\Msun$ of gas to be accreted onto the central black hole at a rate of $3-5\Msunpyr$. At the same time, far 
UV imaging has revealed no trace of star formation or emission from a nuclear point source \citep{mcnamara09}. 
Therefore, a surprisingly large fraction of the $<3\e{9}\Msun$ molecular gas supply \citep{salome08} would need to be
consumed over roughly $100\Myr$ in order to power the outburst. Unless MS0735 hosts an ultramassive black hole with a
mass approaching $10^{11}\Msun$, it would be difficult to power the AGN by Bondi accretion \citep{mcnamara09}. Tapping
the spin energy of the black hole would ease the demands on accretion, although this suggestion has its own problems
\citep{mcnamara11}. 

Here we present a deep ($\sim450\ks$) \textit{Chandra} X-ray observation of MS 0735.6+7421. In Section 2 we describe 
the observations and data reduction methods. Projected and deprojected profiles of cluster properties are presented in
Section 3. New, more accurate measurements for cavity and shock energetics are presented in Section 4. In Section 5 we
present evidence for a rejuvenated AGN outburst and provide estimates of its energetics. In Section 6 we present the
first comparison of outburst interval to central cooling time and cavity heating time for the $11$ objects that have
measured outburst intervals. All results are summarized in Section 7. Throughout this analysis we assume a flat 
$\Lambda \rm{CDM}$ cosmology with $H_0=70\kmpspMpc$, $\Omega_{\rm m}=0.3$, and $\Omega_{\Lambda}=0.7$. This gives an
angular scale of $1$ arcsec $=3.5\kpc$ at the redshift of MS0735, $z=0.216$. All errors are $1\sigma$ unless otherwise
stated.

\begin{table}
\caption{Exposure Times.}
\begin{center}
\begin{tabular}{l c c c}
\hline
Obs ID & Date & Exposure Time & Cleaned \\
 & & (ks) & Exposure Time (ks) \\
\hline
10468 & June 21, 2009 & 46.0 & 41.7 \\
10469 & June 11, 2009 & 93.3 & 86.7 \\
10470 & June 16, 2009 & 142  & 134.8 \\
10471 & June 25, 2009 & 19.5 & 18.1 \\
10822 & June 18, 2009 & 75.4 & 71.8 \\
10918 & June 13, 2009 & 65.2 & 61.8 \\
10922 & June 26, 2009 & 35.4 & 32.3 \\
\hline
Total & & 477 & 447 \\
\hline
\end{tabular}
\end{center}
\label{tab:exposure}
\end{table}

\begin{figure}
\centering
\includegraphics[width=\columnwidth]{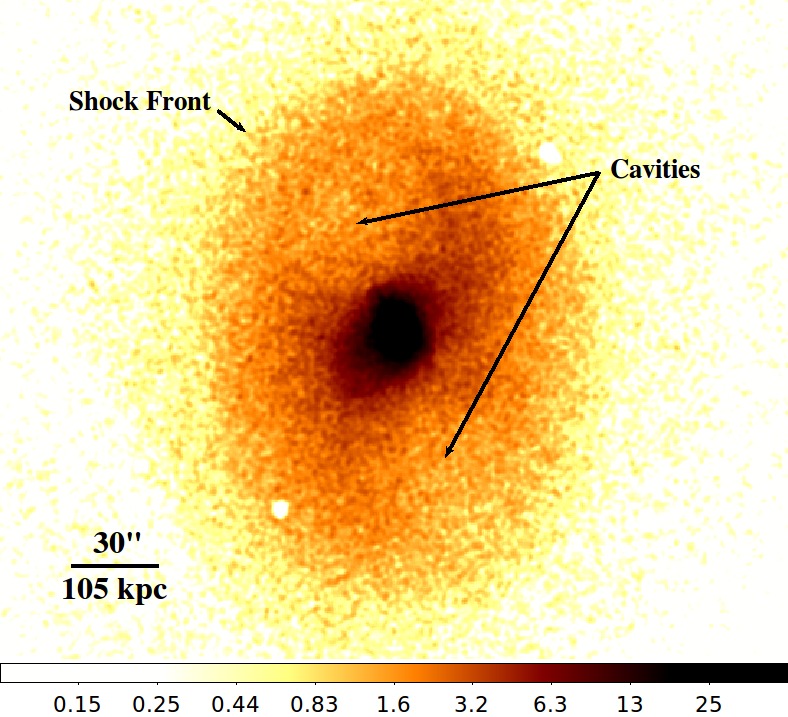}
\caption{\textit{Chandra} X-ray image ($0.5-7\keV$) of MS 0735.6+7421 in units of counts pixel$^{-1}$ and Gaussian-smoothed with a $3$ pixel ($1.5''$) kernel radius. In this image North is up and East is to the left. 
Two cavities and a shock front are visible. Point sources have been excluded from the image.
}
\label{fig:ms07}
\end{figure}

\section{Observations and Data Reduction}
\label{sec:data}

This analysis combines seven \textit{Chandra} observations of MS0735.6+7421 taken on the ACIS-I detector in June 2009. 
The cumulative exposure time for these observations is $477\ks$. Table \ref{tab:exposure} summarizes the exposure times
for each observation. Each observation was reprocessed with \textsc{CIAO} version 4.5 and \textsc{CALDB} version 4.5.6,
which were provided by the \textit{Chandra} X-ray Center. The level 1 event files were reprocessed to apply the latest
gain and charge transfer inefficiency correction and then filtered to remove photons detected with bad grades. The
additional data obtained in VFAINT mode were used to improve screening of the particle background. Background light
curves were extracted from the level 2 events files of chip 0 on the ACIS-I detector. These background light curves were
filtered using the \textsc{lc\_clean} script provided by M. Markevitch in order to identify periods affected by flares.
None of the observations showed any significant flares. The final cleaned exposure time was $447\ks$.

The cleaned events files were then reprojected to match the position of the observation with obs ID 10468. An image for
each observation was produced by summing events in the energy range $0.5-7.0\keV$. These images were then summed to
create a single image for identifying features in the X-ray emission. Point sources were identified using 
\textsc{wavdetect} \citep{wavdetect}. The identified point sources were inspected visually and excluded from subsequent
analysis. The final image, with point sources removed, is shown in Figure \ref{fig:ms07}. The image is not corrected for
exposure.

Blank-sky backgrounds were extracted for each observation, processed the same way as the events files, and reprojected 
to the corresponding position on the sky. Each blank-sky background was normalized to the $9.5-12\keV$ energy band in 
the observed data set. This was a $10-15\%$ correction for all observations. The normalized blank-sky background data 
sets were compared to source-free regions of the observed data set for consistency, and were found to be a close match 
to the background of the observed data set.

Spectral data was analyzed by first extracting source spectra and background spectra from each observation independently.
All extracted spectra and background spectra were summed and the exposure times were adjusted accordingly. Spectra may 
be summed because each observation was taken on the same detector over the course of about $2$ weeks. The roll angles 
are similar, so spectral extraction regions are from similar regions of the chip and therefore have similar responses.
Auxiliary response files were weighted by the number of counts in the spectrum and summed using the \textsc{addarf}
command. The redistribution matrix files were also weighted by the number of counts and summed using \textsc{addrmf}.
Finally, the summed spectrum was binned to a minimum of 20 counts per energy bin.

The loss of area resulting from point sources and chip gaps was corrected by creating an exposure map for each 
observation with the \textsc{mkexpmap} command in \textsc{ciao}, omitting effective area and quantum efficiency. Each
exposure map was weighted by exposure time, summed, and then normalized by the total exposure time. The appropriate area
correction was obtained from the mean value of this exposure map within the region of interest. The correction was
applied to the spectrum and background spectrum by setting the \textsc{areascal} keyword.

\begin{figure}
\centering
\includegraphics[width=0.99\columnwidth]{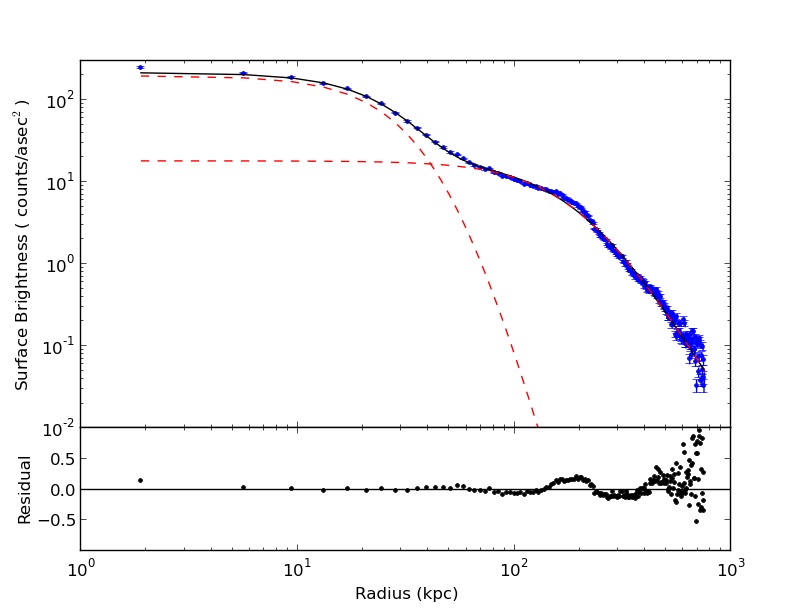}
\caption{Elliptical surface brightness profile fit by a double-$\beta$ model. The blue points show the observed surface brightness profile, while the solid black line is the best fit obtained from the double-$\beta$ model. The dashed red lines show the individual components of the double-$\beta$ model. The residuals are normalized by model values and are shown in the bottom panel.}
\label{fig:betafit}
\end{figure}

\begin{figure}
\centering
\includegraphics[width=\columnwidth]{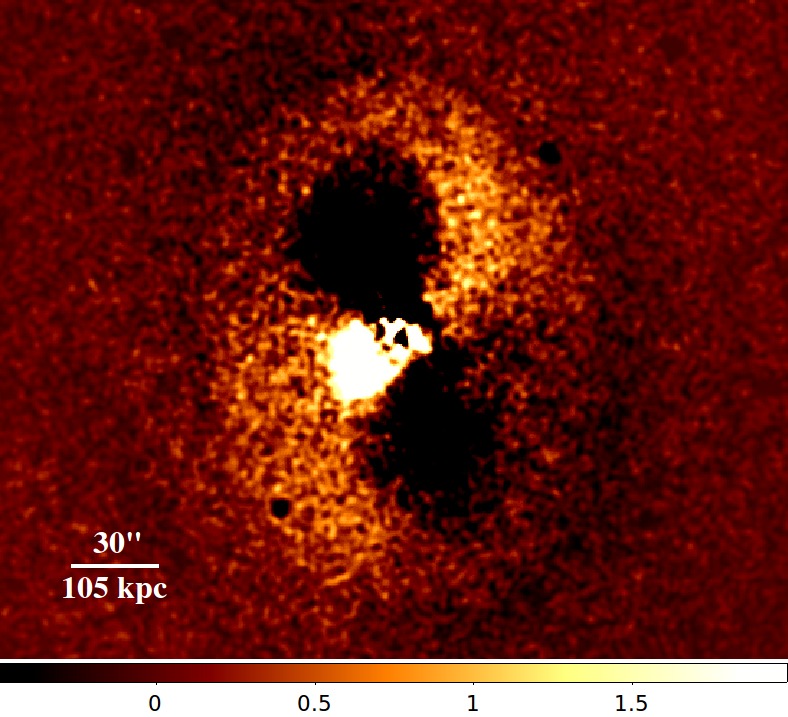}
\caption{Residual image after subtracting a double-$\beta$ model from the X-ray image. The image is in units of counts pixel$^{-1}$ and is Gaussian-smoothed with a $2''$ kernel radius. The dark regions to the NE and SW correspond to the two large cavities in MS0735. A surface brightness edge surrounds these cavities and corresponds to a weak shock front.}
\label{fig:betasub}
\end{figure}

\begin{figure*}
\centering
\begin{minipage}{\textwidth}
\includegraphics[width=0.33\textwidth]{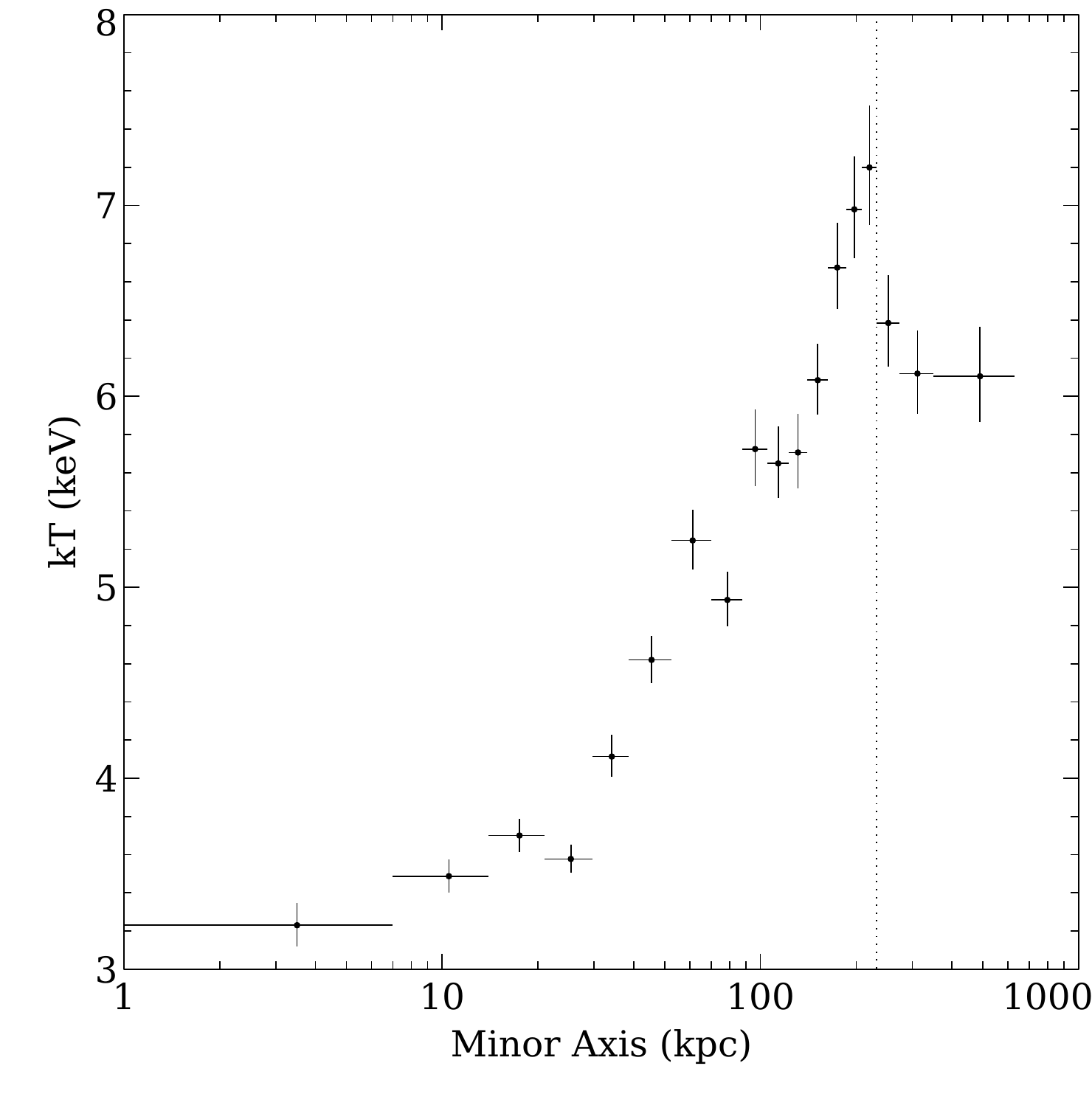}
\includegraphics[width=0.33\textwidth]{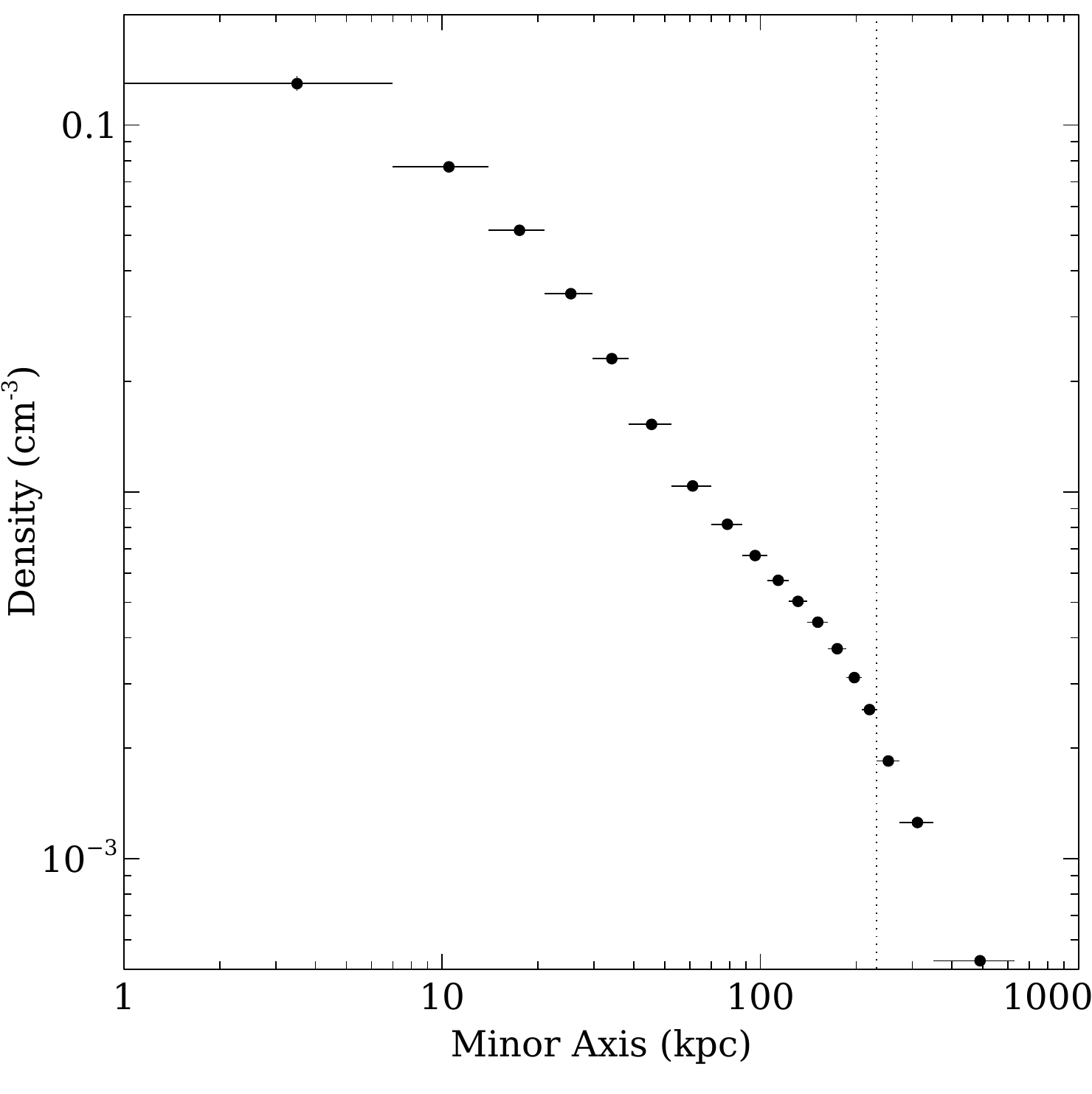}
\includegraphics[width=0.33\textwidth]{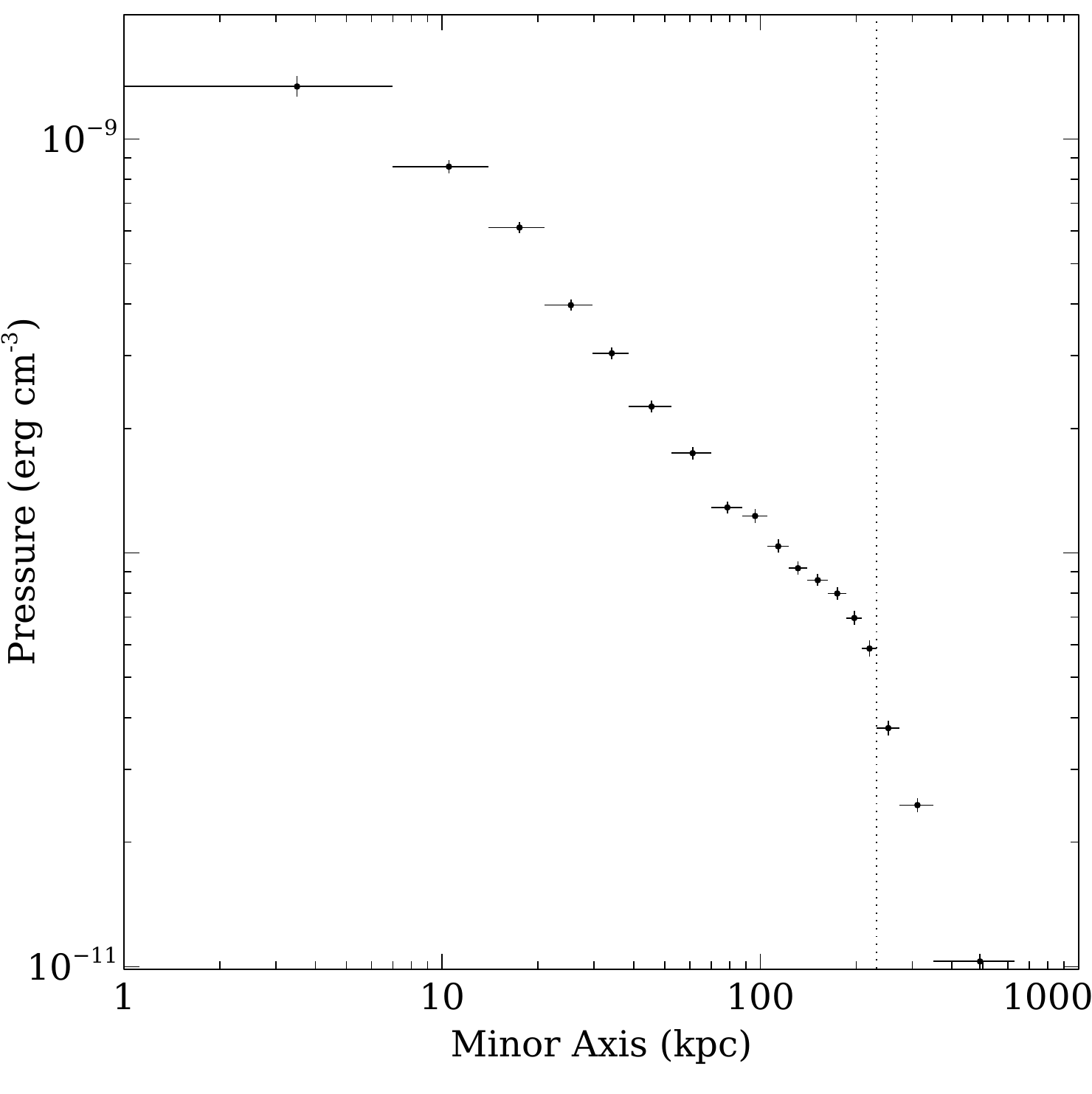}
\end{minipage}
\begin{minipage}{\textwidth}
\includegraphics[width=0.33\textwidth]{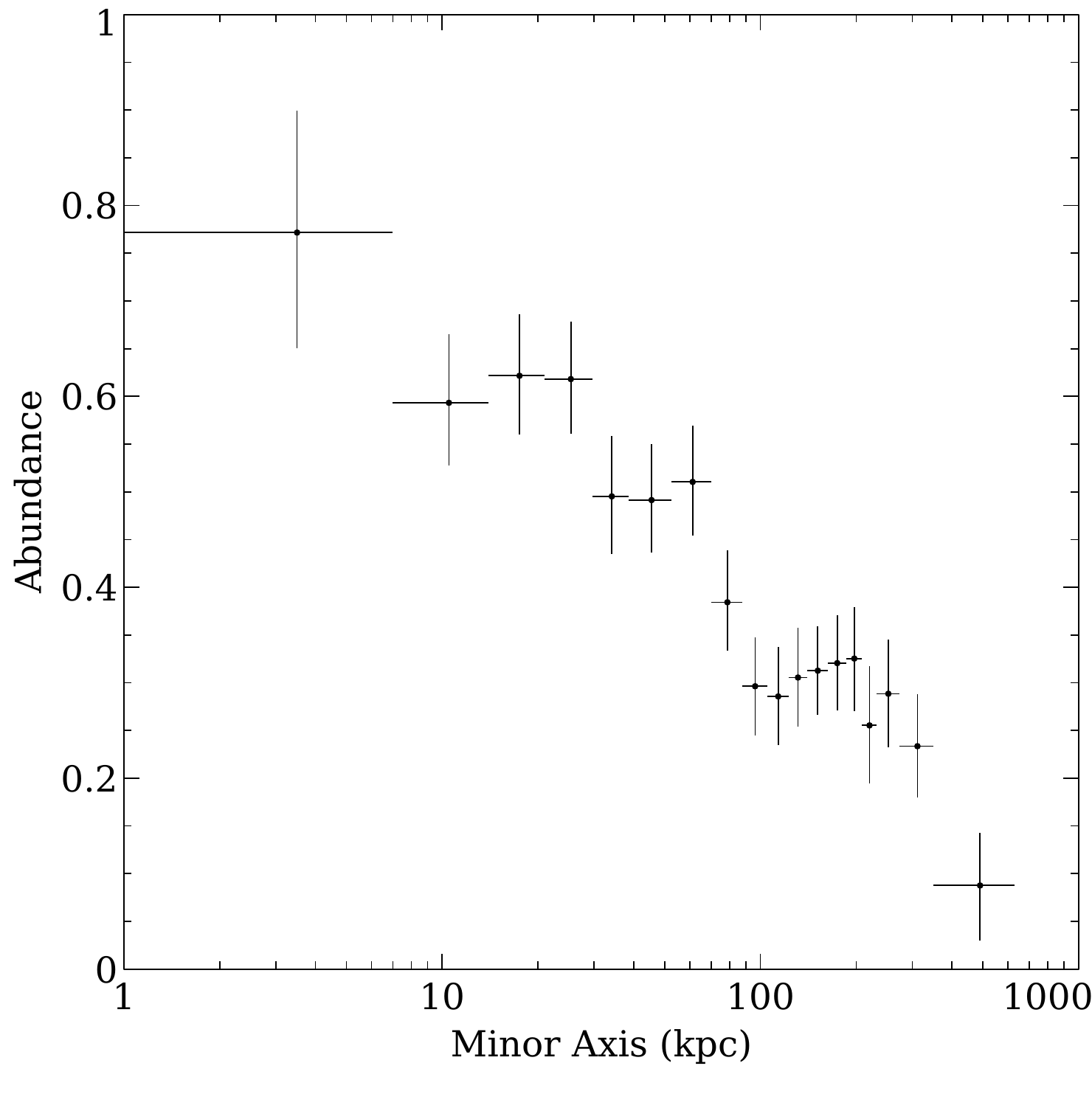}
\includegraphics[width=0.33\textwidth]{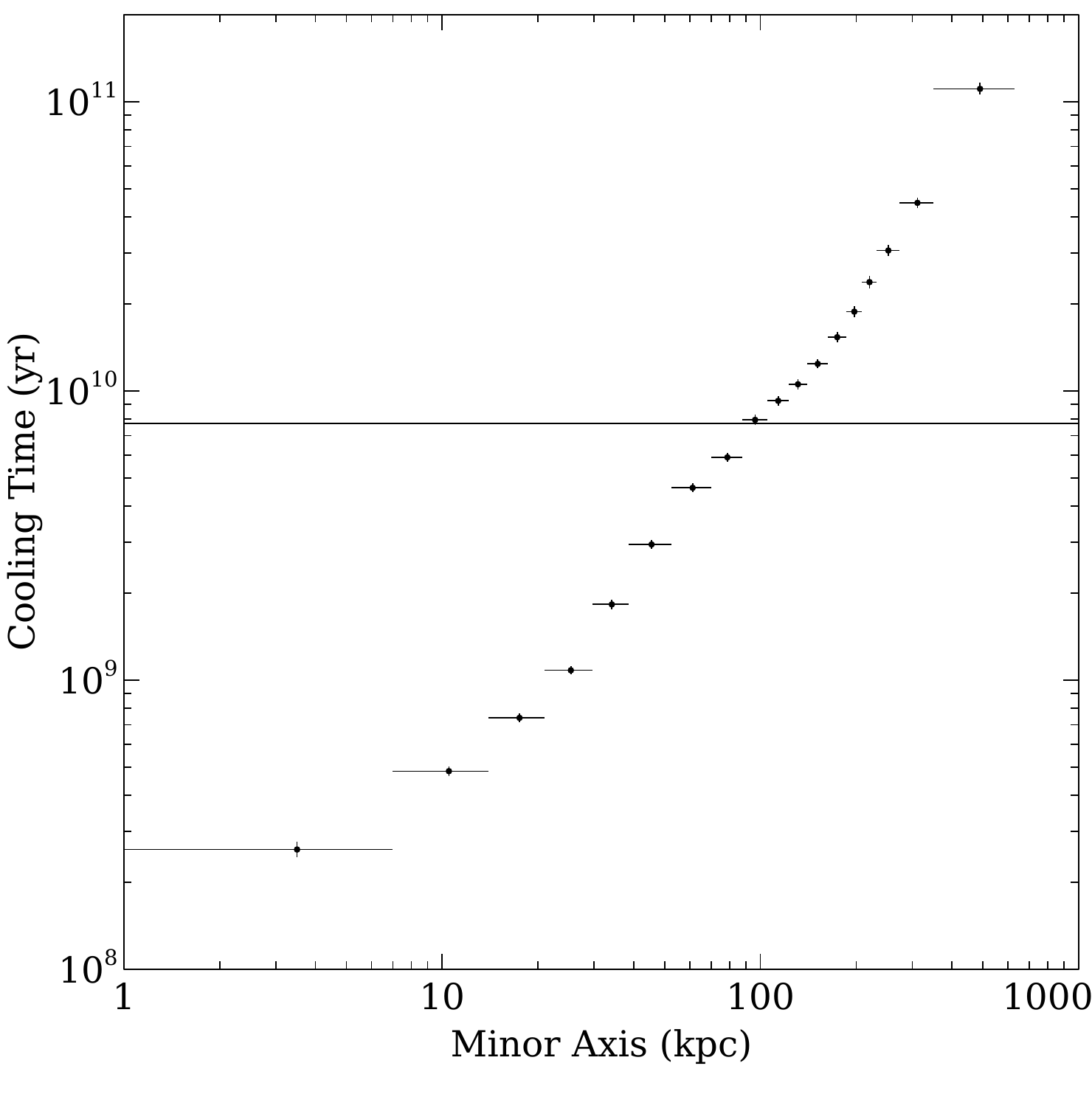}
\includegraphics[width=0.33\textwidth]{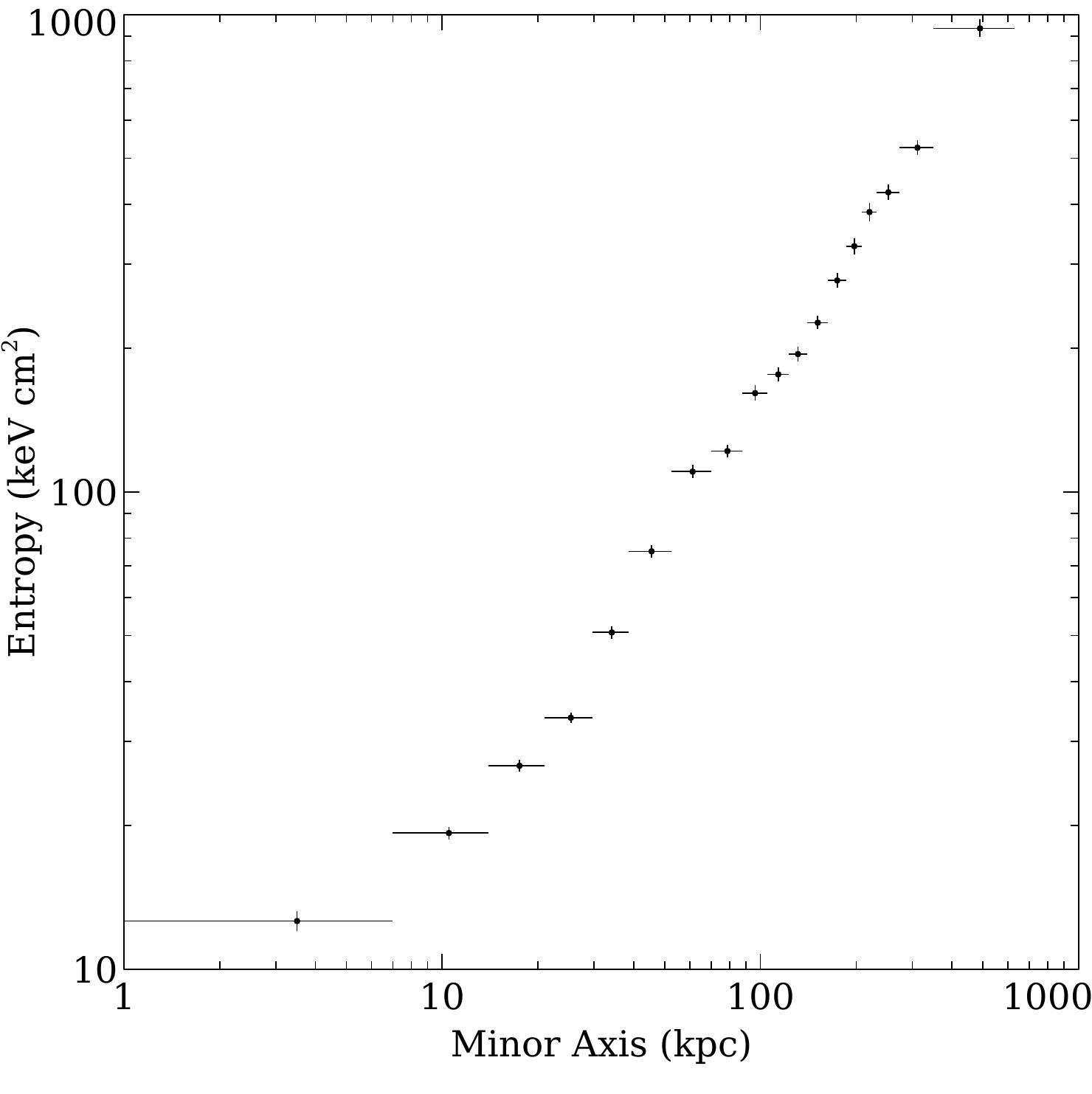}
\end{minipage}
\caption{
Projected temperature, density, pressure, abundance, cooling time, and entropy profiles. The profiles were created using elliptical annuli that are concentric with the outer shock front. All values are plotted against the semi-minor axis of the elliptical annulus. The dotted line in the temperature, density, and pressure profiles shows the approximate location of the outer shock front. Projected measurements overestimate the central density by up to a factor of 2 within $\sim30\kpc$, but are accurate toward larger radii. Refer to the deprojected profiles in Figure \ref{fig:deproj} for accurate central densities.
}
\label{fig:profiles}
\end{figure*}

\section{Cluster X-ray Properties}
\label{sec:xrays}

In this section we present an analysis of the X-ray properties of MS0735, including profiles of surface brightness,
temperature, and abundance as well as quantities derived from these profiles. Maps of temperature and abundance are also
presented. We find that MS0735 is a relaxed, cool core cluster with no evidence of a recent merger. This system hosts 
the largest known AGN outbursts, with cavities with diameters approaching $200 \kpc$ and a cocoon shock enveloping the
cavities. These features are discussed in turn in Section \ref{sec:bigoutburst}.

\subsection{Surface Brightness}

Cluster surface brightness was extracted from the X-ray image for a series of concentric elliptical annuli. The annuli 
had a spacing of $1''$ ($3.5 \kpc$) along the minor axis and were concentric with the shock front. The resulting surface
brightness profile, after background subtraction, is shown in Figure \ref{fig:betafit}. An isothermal, single-$\beta$
model \citep{cavaliere76}, described by
\begin{equation} I_X = I_0 \left[ 1+ \left( R/R_c \right)^2 \right]^{-3\beta+\frac{1}{2}}, \end{equation}
is a poor fit to this profile. Instead, we fit the surface brightness profile with a double-$\beta$ model to account for
the excess emission from the cool core of the cluster. The core was best fit by a $\beta$-model with a normalization of
$200$ counts/arcsec$^2$, a scale radius of $50 \kpc$, and a $\beta$ of $1.8$. The normalization of the second component
was $20$ counts/arcsec$^2$, with $R_{c_2}=200 \kpc$ and $\beta_2=0.9$. The double-$\beta$ model is a poor fit beyond 
$\sim200\kpc$ because of the shock front at $245\kpc$, but is a reasonable fit in the cluster centre.

Subtracting the best-fitting double-$\beta$ model from the cluster emission produces the image shown in Figure 
\ref{fig:betasub}. Two large cavities are easily visible in this image. Sharp edges surround the northeastern (NE) 
cavity, causing it to be more well-defined than the southwestern (SW) cavity. Bright emission is located at the same
radius as the cavities, and could be produced from gas displaced by the cavities. The knot of bright emission 
$\sim20''$ east of the centre corresponds to cool, extended emission. The cocoon shock enveloping the cavities is also
evident in Figure \ref{fig:betasub}. The large positive residual between $\sim150-250\kpc$ is located along the inner
edge of the shock front ($245\kpc$) and is caused by gas that has presumably been displaced by the cavities.

\subsection{Projected Profiles}
\label{sec:profiles}

Radial profiles of temperature, density, and abundance were created by fitting spectra extracted from elliptical annuli.
Each elliptical annulus was taken to have the same major-to-minor axis ratio ($1.37$) and position angle ($7^{\circ}$
east of north) as the outer shock front, which was determined by eye. The volume of each annulus was calculated assuming
prolate symmetry.
The inner $6$ annuli were created with a fixed radial bin width of $7 \kpc$ along the minor axis, which was chosen based
on the $1''$ spatial resolution of \textit{Chandra}. These regions contained between $4000$ and $10000$ net counts. The
remaining annuli were created with the number of counts increasing with radius, with the outermost annulus containing 
$\sim 30000$ net counts. Projected temperature, density, pressure, abundance, cooling time, and entropy profiles are 
shown in Figure \ref{fig:profiles} and are plotted against the semi-minor axis of the annulus.

\subsubsection{Spectral Fitting}
\label{sec:model}

Projected gas properties were determined by fitting each extracted spectrum with an absorbed single-temperature 
\textsc{phabs(mekal)} model \citep{mekal1,mekal2,mekal3,mekal4} in \textsc{xspec 12} \citep{arnaud96}. Temperature,
normalization, and abundance were allowed to vary while the foreground column density, $N_H$, was fixed to the Galactic
value of $3.1\e{20}\,\rm{atoms}\psqcm$ \citep{kalberla05,hartman97}. This value is consistent with fitted values in the
outskirts of the cluster within $1\sigma$. Abundance line ratios were set to the values given in \citet{angr} for
consistency with previous work.

Density is related to the normalization of the \textsc{mekal} thermal model through the emission integral 
$\int{n_e n_H\, \rm{dV}}$. The electron and hydrogen densities, $n_e$ and $n_H$, are taken to be constant 
within each annulus. All emission in an annulus is assumed to originate from a prolate ellipsoid. 
Assuming hydrogen and helium mass fractions of $\rm{X}=0.75$ and $\rm{Y}=0.24$ \citep{angr} gives 
$n_e=1.2\, n_H$.  Density can then be determined from the normalization of the thermal model and the volume 
of the ellipsoid. Pressure is determined from temperature and density using the ideal gas law, $P=2n_e kT$.

\begin{figure*}
\centering
\begin{minipage}{0.95\textwidth}
\includegraphics[width=0.49\columnwidth]{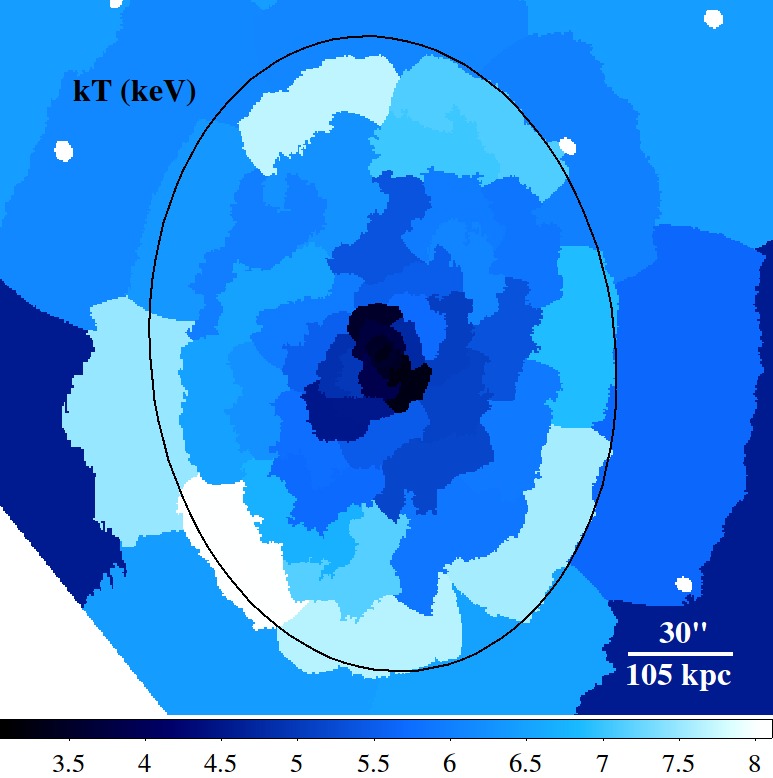}
\includegraphics[width=0.49\columnwidth]{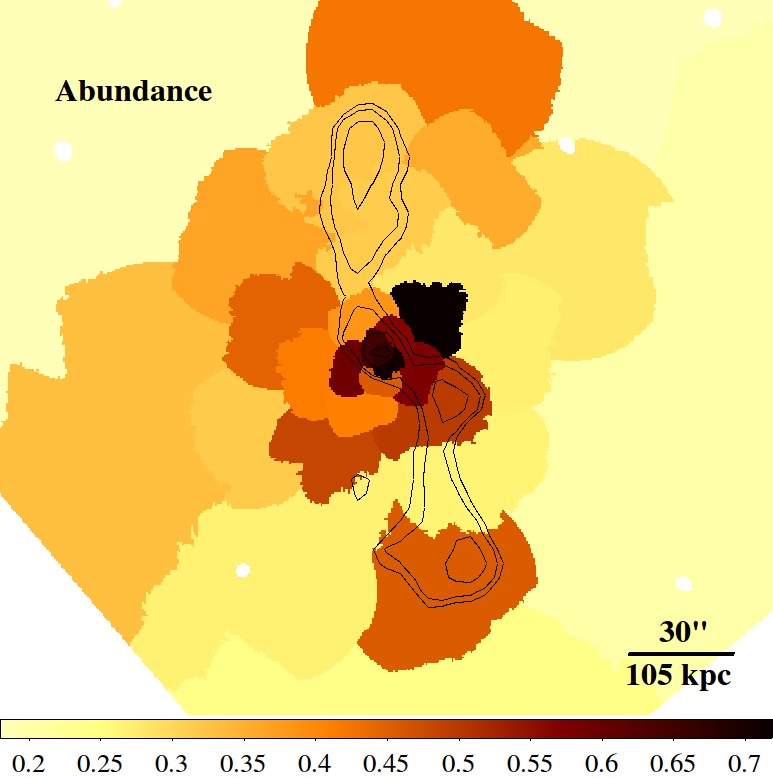}
\end{minipage}
\caption{Temperature (left) and abundance (right) maps created by grouping regions of similar surface brightness. The colourbars give the temperature and abundance in units of $\keV$ and $Z_{\odot}$, respectively. Typical errors are $5\%$ for temperature and $15-20\%$ for abundance. Point sources have been excluded from the images. The black ellipse in the temperature map outlines the qualitative fit to the shock front as determined from the $0.5-7.0 \keV$ X-ray image. A temperature jump is visible across the shock front. $327 \MHz$ radio contours are overlaid on the abundance map.
}
\label{fig:maps}
\end{figure*}

\subsubsection{Temperature Profile}

The observed temperature profile, shown in the upper left panel of Figure \ref{fig:profiles}, is consistent with 
the profiles from \citet{vikh05}. The mean projected temperature within $10 \kpc$ of the cluster centre is 
$3.23\pm0.11\keV$. It rises to a maximum of $7.2\pm0.3\keV$ at $220 \kpc$, where it drops abruptly to $6.4\pm0.2\keV$. 
The temperature beyond this radius is roughly constant at $\sim6.1\keV$. The sharp drop in temperature at $220 \kpc$ 
is interpreted as a weak shock front (see \citealt{nature05}). The ratio between post- and preshock temperatures, 
$1.13\pm0.06$, is consistent with the projected temperature jump of $\sim10\%$ expected from the shock measurement. 
This shock, with a Mach number $1.26$ determined from the surface brightness profile, is discussed further in 
Section \ref{sec:outer_shock}.

The decrease in temperature at $25\kpc$ corresponds to cool gas that extends to the southeast, which coincides with the
bright residual from Figure \ref{fig:betasub}. This feature is evident in the temperature map discussed in Section 
\ref{sec:maps}. Significant scatter about the mean temperature is observed at $\sim80\kpc$, resulting from excess 
emission that extends perpendicular to the jet axis. This emission is visible in Figure \ref{fig:betasub}, and may
correspond to gas that has been displaced by the cavities.

\subsubsection{Abundance Profile}

The metal abundance within $10\kpc$ of the cluster centre is $0.77\pm0.12\Zsun$, and then decreases with radius. The
abundance rise toward the cluster centre is likely due to enrichment from supernovae in the BCG \citep{degrandi09}. 
From $\sim100-300\kpc$ the abundance flattens to $0.3\Zsun$, which is typical of cool core clusters 
\citep{degrandi01,degrandi04}. The last radial bin has an abundance of $\rm{Z}=0.09\err{0.06}{0.05}\Zsun$, 
$3\sigma$ below the $0.3\Zsun$ plateau.

\subsubsection{Entropy Profile}

The entropy index of intracluster gas, defined as $K=kT\, n_e^{-2/3}$, offers a more direct insight into heating and
cooling processes than either temperature or density individually. Above $\sim20\kpc$ the entropy index can be fit by a
powerlaw with slope $1.07$, which is consistent with the slopes found by \citet{voit05}. Toward the centre the mean
entropy index appears to flatten toward the central projected value of $12.6\pm0.6\keV\cmsq$. However, recent work from
\citet{panagoulia13} has shown that this may be a resolution effect and that entropy should continue to decrease toward
the cluster centre.

\subsubsection{Cooling Time}
\label{sec:Lx}

If uncompensated by heating the ICM will radiate away its thermal energy on a timescale 
$t_{cool}=3p/ \left[ 2n_en_H\Lambda(Z,T) \right]$, where $\Lambda(Z,T)$ is the cooling function as a 
function of metallicity, $Z$, and temperature, $T$. The cooling function is determined from the X-ray 
bolometric luminosity, which is given by $L_{\rm{X}}=\int n_en_H \Lambda(Z,T)\, \rm{dV}$ and is obtained 
by integrating the unabsorbed thermal model between $0.1$ and $100\keV$. The cooling time in MS0735, 
shown in Figure \ref{fig:profiles}, roughly follows a powerlaw with a slope of $1.4$. The projected 
cooling time within the central $10\kpc$ is $(2.6\pm0.2)\e8\yr$. Projection effects have a significant 
effect on central density, and therefore on central cooling time. The deprojected profile is presented 
in Section \ref{sec:deproj}.

\begin{figure}
\centering
\includegraphics[width=\columnwidth]{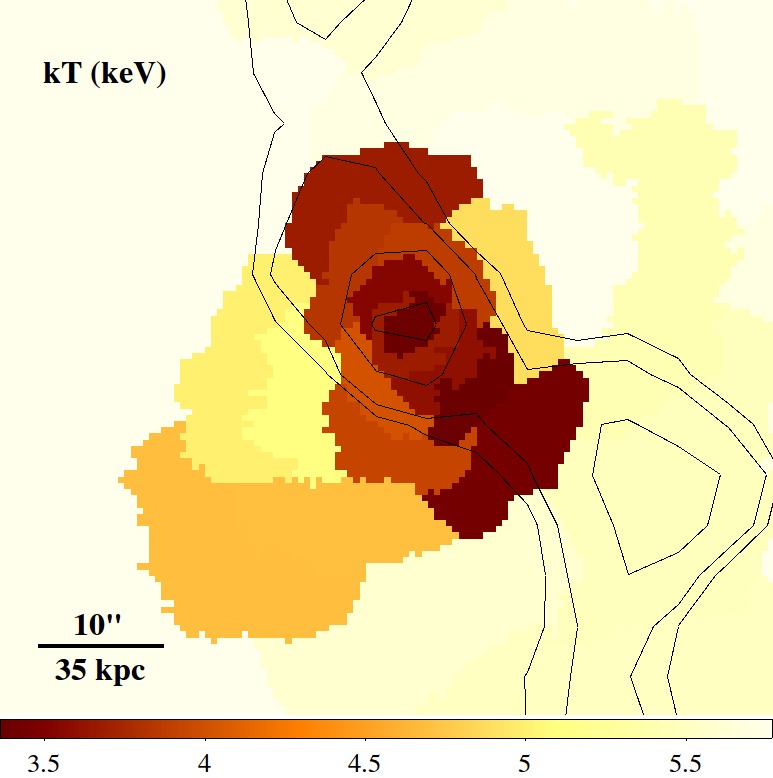}
\caption{Inner $200 \kpc$ of the temperature map presented in Figure \ref{fig:maps}. Low temperature gas from the centre of the cluster is entrained in the radio jet (black contours).}
\label{fig:entrainment}
\end{figure}

\subsection{Maps}
\label{sec:maps}

Local variations in cluster abundance and temperature are traced using maps created with the \textsc{contbin}
\footnote{http://www-xray.ast.cam.ac.uk/papers/contbin} method \citep{contbin}. Cluster emission is grouped into bins 
that closely follow surface brightness. The minimum signal-to-noise was set to $60$ ($3600$ net counts) for the
temperature map in order to produce an image with high spatial resolution. Accurate abundance measurements require more
counts, so a signal-to-noise of $80$ ($6400$ net counts) was used for this map. A spectrum was extracted from each region
produced by \textsc{contbin} and was fit with the absorbed thermal model described in Section \ref{sec:model}. The maps
are shown in Figure \ref{fig:maps} and are both in broad agreement with the profiles in Figure \ref{fig:profiles}. Point
sources have been excluded from these maps.

\subsubsection{Temperature Map}

The approximate location of the outer shock front is indicated by the black ellipse overlaid on the 
temperature map (Figure \ref{fig:maps}, left). A clear temperature jump is seen between the pre- and 
post-shock bins along the entire shock front. The ratio of postshock to preshock temperature varies 
between $1.15$ and $1.3$ along the shock front. Accounting for projection effects, these jumps slightly 
exceed expected jump for the Mach $1.26$ shock reported in this work. See Section \ref{sec:outer_shock} 
for a more detailed analysis.

An extended region of cooler emission is seen to the immediate SE of the cluster core, coinciding with 
the bright emission seen in Figure \ref{fig:ms07} (right). The temperature of this gas, $4.5\pm0.2\keV$, 
is cooler than the surrounding $\sim5.5\keV$ gas. The drop in temperature observed in Figure \ref{fig:profiles} 
is consistent with the location of this extended feature.

The central $\sim200\kpc$ of the temperature map is shown in Figure \ref{fig:entrainment} and is overlaid 
with $327\MHz$ VLA radio contours \citep{birzan08}. Cooler gas ($\sim3.5\keV$), which likely originates 
from the centre of the cluster, extends along the direction of the radio jets, implying that the ICM is 
entrained in the radio jet and is being dragged to high altitudes. This effect has also been observed 
in Hydra A, where the energy required to uplift the cool gas is comparable to the work required to inflate 
the cavities \citep{gitti11}. Removing the supply of low entropy gas near the cluster centre slows the 
rate of cooling, which is an important step in regulating AGN feedback \citep{clif09}.

\subsubsection{Abundance Map}

The central abundance in MS0735 is roughly $0.7\Zsun$ and decreases with radius until it reaches 
$0.2-0.3\Zsun$. Regions of enhanced abundance are seen toward the end of the radio jets
(black contours). A region with $Z=0.46\Zsun$ lies at the end of the jet pointing to the south. To 
the north the $0.42\Zsun$ region lies beyond the extent of the radio jet. This observation is consistent 
with work done by \citet{clif09} and \citet{simionescu09}, who argued that cavities may lift metals to 
large radii as they rise through the cluster. Kirkpatrick et al. (in prep) measure excess iron emission 
along the direction of the radio jet in MS0735 out to a radius of $\sim300\kpc$. The high metallicity 
regions in Figure \ref{fig:maps} are located at distances consistent with this iron radius.

\begin{figure*}
\centering
\begin{minipage}{0.95\textwidth}
\centering
\includegraphics[width=0.4\columnwidth]{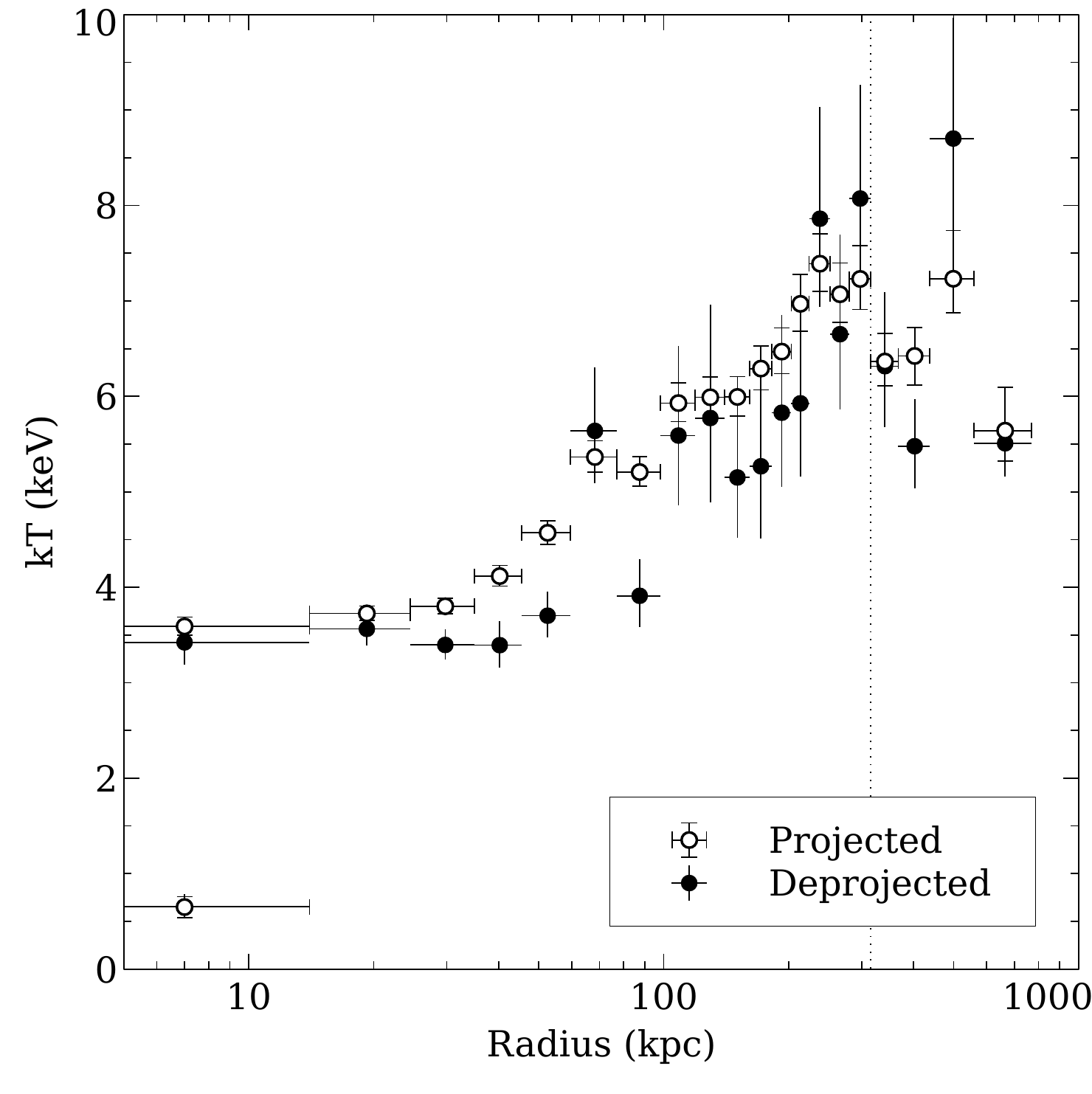}
\includegraphics[width=0.4\columnwidth]{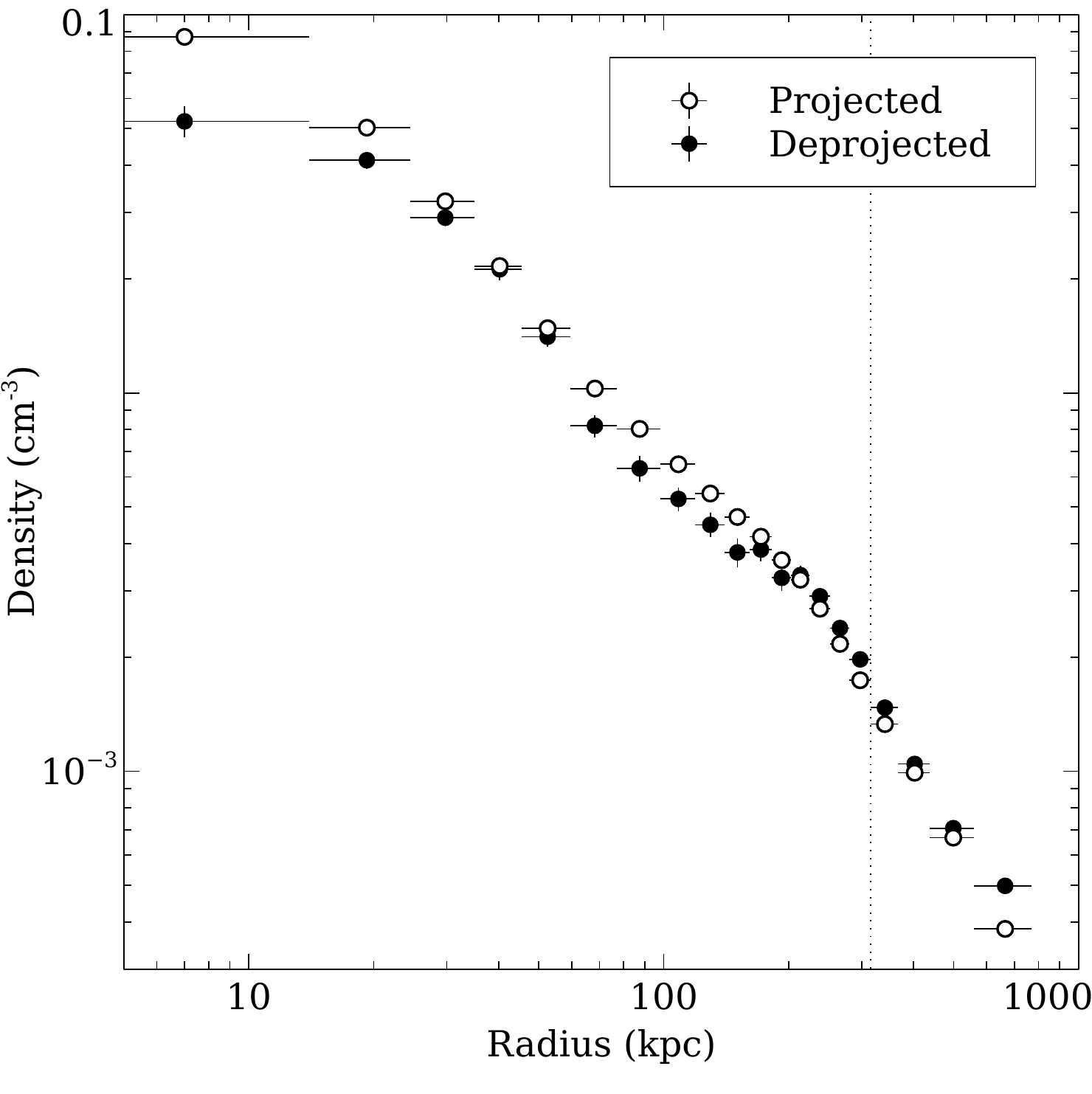}
\end{minipage}
\begin{minipage}{0.95\textwidth}
\centering
\includegraphics[width=0.4\columnwidth]{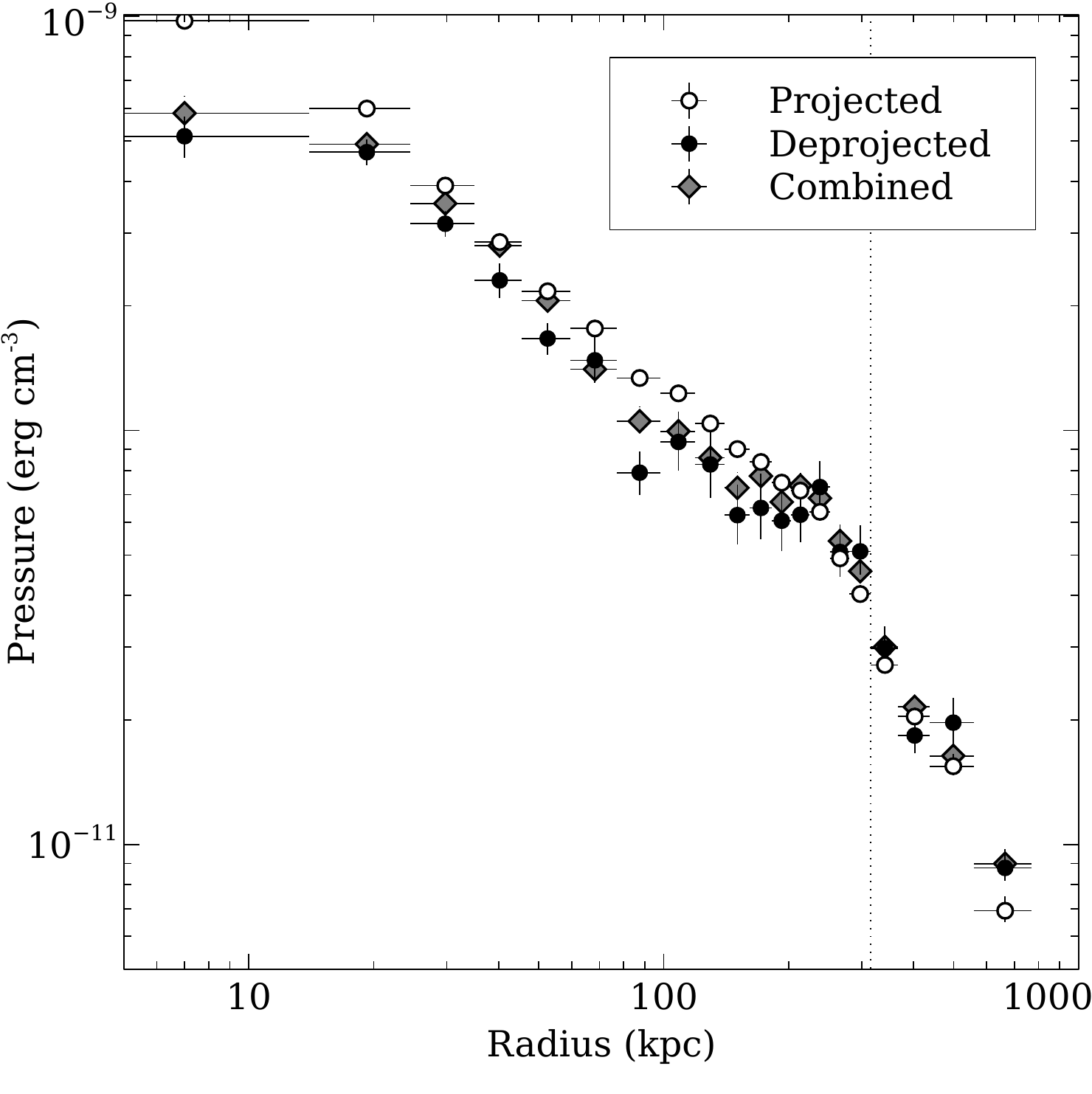}
\includegraphics[width=0.4\columnwidth]{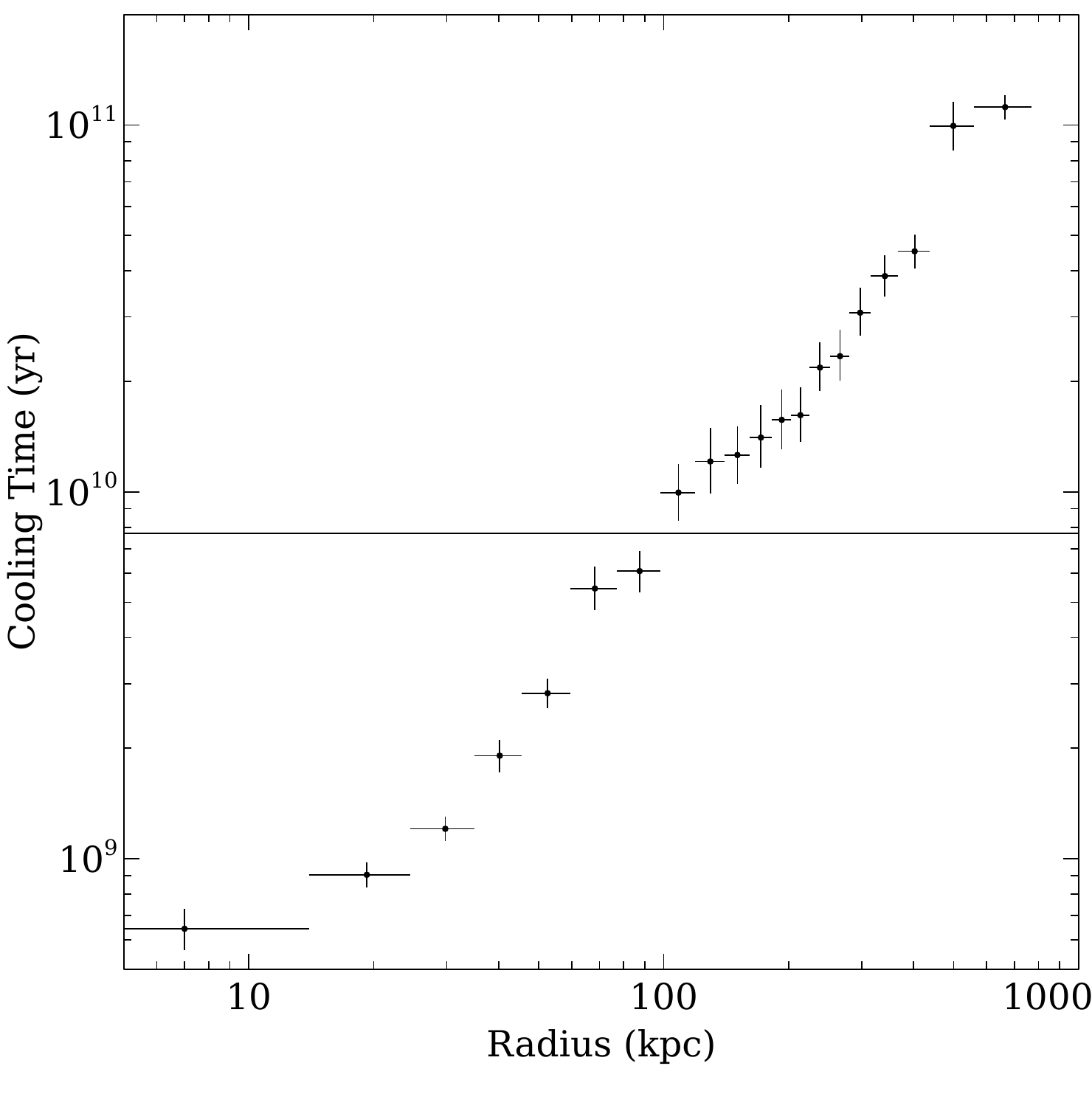}
\end{minipage}
\caption{Projected (open circles) and deprojected (filled circles) profiles of temperature, density, and pressure. The dotted line shows the location of the temperature jump, which is consistent with the location of the shock. Deprojection was performed using the model-independent \textsc{dsdeproj}. Including a second temperature component in the central region was found to improve the fit. The gray diamonds in the pressure profile (lower-left) correspond to pressure values calculated using deprojected densities and projected temperatures. The cooling time, determined from the deprojected profiles, is shown in the lower-right panel.}
\label{fig:deproj}
\end{figure*}

\subsection{Deprojected Profiles}
\label{sec:deproj}

Projected foreground and background emission skews central projected densities and temperatures to higher 
values. We now correct for this effect by employing deprojected fits to radial profiles. Spectra were 
extracted from concentric circular annuli containing roughly $13000$ net counts per annulus. These spectra 
were area-corrected, then deprojected using the model-independent \textsc{dsdeproj} package described 
in \citet{dsdeproj1} and \citet{dsdeproj2}. The absorbed thermal model described in Section \ref{sec:model} 
was fit to each of the deprojected spectra, and the resulting profiles are shown in Figure \ref{fig:deproj}. 
The ``onion-peeling'' approach of the \textsc{projct} model produces fits that are consistent with \textsc{dsdeproj}.
Projected profiles were created using these regions and are plotted alongside the deprojected profiles for 
comparsion. Two data points are plotted in the central bin due to the presence of multiphase gas, which is 
discussed in detail in Section \ref{sec:multiphase}.

Deprojection subtracts the line-of-sight contribution of overlying gas from each annulus. The resulting 
central density, $(5.2\pm0.5)\e{-2}\pcmcu$, is therefore significantly smaller than the central projected 
density. The deprojected central temperature, on the other hand, is consistent with the projected temperature 
within $1\sigma$. The temperature of the cold phase in the central region is consistent between projected 
and deprojected profiles.

The projected temperature profile increases smoothly between $20$ and $100\kpc$, while the deprojected 
profile varies significantly. The deprojection relies critically on the assumption of spherical symmetry, 
which clearly fails here and contributes to the bouncing deprojected temperatures. The drop in projected 
temperature seen at this location in Figure \ref{fig:profiles} is exaggerated by the deprojection. The 
pressure profile used for subsequent calculations of cavity and shock properties combines deprojected densities 
with projected temperature in order to avoid the unstable solutions in the deprojected temperature profile.

At a radius of $315\kpc$ the projected temperature jumps from a preshock value of 
$6.4\pm0.3\keV$ to a postshock value of $7.2\pm0.3\keV$. This temperature 
jump, $1.14\pm0.07$, is consistent with the jump observed in Figure \ref{fig:profiles}. The deprojected 
profile shows a larger jump, $1.3\pm0.2$, but it is consistent with the projected jump within $1\sigma$. 
The location of this jump is consistent with the location of the shock front, which extends from $245\kpc$ 
along its minor axis to $335\kpc$ along its major axis.

\subsection{Multiphase Gas}
\label{sec:multiphase}

\begin{figure}
\includegraphics[width=0.97\columnwidth]{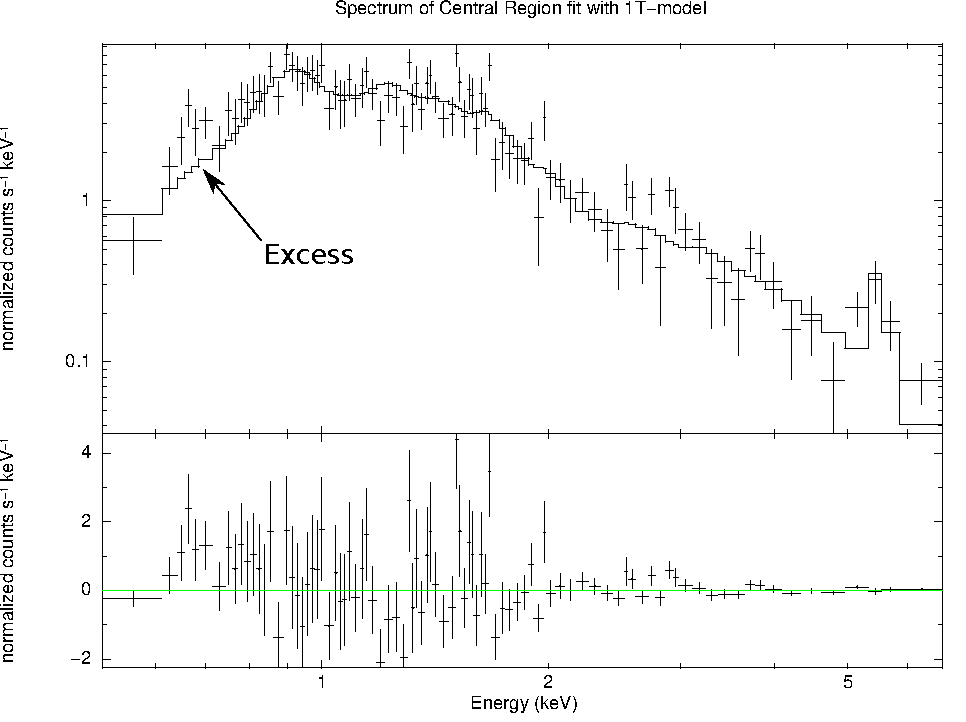}
\caption{Absorbed, single-temperature model fit to the spectrum of the central $14\kpc$. The model is a poor fit to the spectrum at low energies, with a clear excess between $0.6$ and $0.7\keV$. Including a second temperature component in the model improves the fit significantly.}
\label{fig:spectrum}
\end{figure}

A single temperature thermal model provides a poor fit to the spectrum of the central $14\kpc$ below 
$1\keV$. The spectrum, shown in Figure \ref{fig:spectrum}, has a soft excess located between $0.6$ 
and $0.7\keV$, resulting in a column density that approaches zero when allowed to vary. This is 
indicative of lower temperature gas located at the centre of the cluster. A second thermal component 
was added to the original model, with abundances tied between the two thermal components and the 
column density fixed to the Galactic value. Applying an F-test, we find the two-temperature model 
provides a statistically significant improvement to the fit. The best-fitting temperatures are 
$3.42\err{0.26}{0.23}\keV$ and $0.65\pm0.13\keV$, and the normalization of the cold component is 
$20$ times smaller than that of the hot component. A single-temperature model is able to model 
the gas adequately at larger radii.

The density ratio between the cold and hot phases, assuming that they are in pressure balance, 
is given by $n_c/n_h=kT_h/kT_c$. With temperatures of $3.4$ and $0.65\keV$, the cold gas would 
be a factor of $5$ more dense than the hot gas. The mass ratio between these components can then 
determined from the normalizations of the \textsc{xspec} thermal models, $K\propto n^2V$. Since 
$M\propto n_eV$, the cold, dense gas is $\frac{M_h}{M_c}=\frac{K_h}{K_c}\frac{n_c}{n_h}\approx100$ times 
less abundant than the hot gas. The ratio between luminosities is obtained directly from the X-ray 
bolometric luminosities of the model components, and is $L_{\rm{X,h}}/L_{\rm{X,c}}\approx10$.

The cooling time of the hot phase within the central $14\kpc$, $6.4\err{0.9}{0.8}\e8\yr$, is 
a factor of $\sim2.5$ longer than the projected cooling time within $10\kpc$, $(2.6\pm0.2)\e{8}\yr$. 
The cold component, with a cooling time of $\sim1.3\e{7}\yr$, cools on a much shorter 
timescale than the hot phase. The implied mass deposition rate for a pure cooling model is 
$26\pm2\Msunpyr$ for the hot phase and $13\pm5\Msunpyr$ for the cold phase. Both of these deposition 
rates greatly exceed the feeble star formation rate of $<0.25\Msunpyr$ \citep{mcnamara09}. While 
cooling gas is apparently able to sustain the AGN, it is not fuelling star formation at an appreciable rate.

The radiative cooling of the central ICM can be suppressed by AGN heating provided cavity power 
exceeds the X-ray luminosity of the gas. The X-ray luminosities of the hot and cold components 
are $(1.33\pm0.06)\e{43}$ and $(1.2\pm0.3)\e{42}\ergps$, respectively. The mean power of the inner 
cavity system, discussed in Section \ref{sec:smalloutburst}, is $5.2\err{2.6}{1.8}\e{44}\ergps$, 
exceeding the combined X-ray luminosity in the central region by a factor of $40$. While this 
provides ample power to suppress cooling, it is not clear how much of the energy is dissipated 
within the central region.

AGN feedback can also suppress radiative cooling by physically removing the supply of low entropy 
gas from the centre of the cluster \citep{gitti11}. X-ray cavities have been shown to couple to the metal-rich 
central gas, dragging it toward large radii as the bubbles rise through the ICM \citep{clif09,simionescu09}. 
The temperature map in Figure \ref{fig:entrainment} shows that cool gas is displaced along the 
direction of the radio jet, implying that gas has been entrained by the radio jet or X-ray cavities 
and is being displaced from the cluster centre. Using abundance profiles along and perpendicular 
to the direction of the cavities, Kirkpatrick et al. (in prep) measure a gas outflow rate of 
$150\pm80\Msunpyr$ being dragged to a radius of $\sim300\kpc$. This large mass outflow rate can 
easily offset the combined $40\Msunpyr$ mass deposition rate.

\subsection{Cooling Region}

We define the region in which the cooling time is less than $7.7\e{9}\yr$, the look-back time to 
$z=1$ for concordance cosmology, as the cooling region. In the absence of major mergers, some energy 
source must make up for the power radiated from within this region in order to suppress cooling in 
the long term. The cooling radius, $r_{\rm cool}$, obtained from Figure \ref{fig:deproj} is 
$100\kpc$, which is consistent with the XMM-Newton analysis of \citet{gitti07}. A spectrum was 
extracted from a region with radius $r_{\rm cool}$ and was deprojected 
using spectra obtained from 3 overlying regions with the same radial width. The total luminosity 
within the cooling region was estimated by fitting this deprojected spectrum with the absorbed 
thermal model discussed in Section \ref{sec:model} and integrating the thermal component between 
$0.1$ and $100\keV$. The unabsorbed X-ray bolometric luminosity within this region is 
$L_{\rm X}(<r_{\rm cool}) = (2.61\pm0.02)\e{44}\ergps$.

The amount of gas that cools out of the ICM is estimated by adding an \textsc{mkcflow} component 
to the thermal model. The low temperature limit of this model was fixed to $0.1 \keV$ in order 
to provide an upper limit on the amount gas cooling to low temperatures. The high temperature 
component was tied to the temperature of the \textsc{mekal} model and allowed to vary. Abundances 
were also tied between models. The resulting cooling luminosity, $L_{\rm{cool}} < 1.2\err{0.5}{0.4}\e{43}\ergps$, 
is a small fraction ($<5\%$) of the total X-ray luminosity, and corresponds to $<22\pm8\Msunpyr$ of 
gas cooling out of the ICM, which is broadly consistent with the $40\pm10\Msunpyr$ reported 
by \citet{gitti07}. With a star formation rate of $<0.25\Msunpyr$ \citep{mcnamara09}, 
only a small portion of gas actually cools out of the ICM and forms stars.

\begin{table*}
\begin{minipage}{\textwidth}
\caption{List of cavity properties.}
\begin{center}
\begin{tabular}{l c c c c c c c c c c}
\hline
Cavity & $a$ & $b$ & R & $p$V & t$_{\rm buoy}$ & t$_{\rm c_s}$ & t$_{\rm refill}$ & t$_{\rm avg}$ & P$_{\rm{cav}}$ & P$_{\rm{cav,tot}}$ \\
       & kpc & kpc & kpc & $10^{59} \erg$ & $10^7 \yr$ & $10^7 \yr$ & $10^7 \yr$ & $10^7 \yr$ & $10^{44} \ergps$ & $10^{44} \ergps$ \\
\hline
Outer NE & 109\err{17}{15}    & 106\err{17}{16}    & 150   & 110\err{60}{40}   & 9.1 & 12  & 24  & 15 & 90\err{50}{35} & \multirow{2}{*}{170\err{60}{50}} \\
Outer SW & 120\err{10}{20}    & 100\err{18}{25}    & 186   & 110$\pm$50        & 11  & 14  & 25  & 17 & 80\err{35}{40}   \\
Inner NE & 13.3\err{1.1}{2.0} & 10.1\err{2.5}{2.1} & 19.3  & 0.9\err{0.5}{0.4} & 3.3 & 1.9 & 7.4 & 4.2 & 2.8\err{1.6}{1.3}  & \multirow{2}{*}{5.2\err{2.6}{1.8}}  \\
Inner SW & 15.5\err{2.3}{2.0} & 10.5\err{3.2}{3.3} & 25.0  & 0.9\err{0.7}{0.5} & 4.1 & 2.5 & 7.8 & 4.8 & 2.4\err{2.0}{1.3}    \\
\hline
\end{tabular}
\end{center}
\label{tab:cavities}
\end{minipage}
\end{table*}

\section{Large Outburst}
\label{sec:bigoutburst}

Two large cavities, with diameters of $\sim200\kpc$, are visible in the X-ray image and are filled by 
radio emission from the relativistic jet. The cavities are surrounded by a continuous surface brightness 
edge, corresponding to a weak but powerful shock front. This section presents new, updated measurements 
of cavity and shock power for the large outburst in MS0735.

\subsection{Cavities}
\label{sec:cavities}

The total energy required to inflate a bubble in the ICM is given by its enthalpy,
\begin{equation} E_{\rm cav} = \frac{\gamma}{\gamma -1} pV, \end{equation}
where $p$ is the cavity's pressure and $V$ is its volume. The ratio of heat capacities, $\gamma$, is 
$4/3$ for a relativistic gas and is $5/3$ for a non-relativistic, monatomic gas. For relativistic 
contents, the total cavity energy is $E_{\rm cav}=4pV$. The age of an outburst is estimated using 
three characteristic timescales: the sound crossing time, buoyancy time, and refill time \citep{birzan04}. 
In general the sound-crossing time is the shortest, refill time is the longest, and buoyancy time lies 
in between. We estimate the age of a cavity by the mean of these three timescales and calculate the 
power required to inflate the cavities as $P_{\rm cav}=4pV/t_{\rm avg}$. This age is likely an overestimate 
of the true age of the bubble, so the powers reported here are underestimates.

The projected sizes and positions of the cavities were determined by eye by fitting ellipses to the 
surface brightness depressions. The cavities are surrounded by bright rims that are presumably composed 
of displaced gas. The midpoint of these rims is taken for the measurement of cavity size. The inner 
(outer) edge of the rim is used to obtain a lower (upper) limit on the projected size of the cavity, 
which is used to determine the errors on cavity volume and subsequently on cavity enthalpy and power.
The northeastern cavity is best fit by an ellipse with semi-major axis $a=109\err{17}{15}\kpc$, semi-minor 
axis $b=106\err{17}{16} \kpc$, and projected distance $R=150 \kpc$ with a position angle of $19.4^{\circ}$ 
east of north. The southwestern cavity is best fit by an ellipse with $a=120\err{10}{20}\kpc$, 
$b=100\err{18}{25}\kpc$, and $R=186\kpc$ with position angle $202^{\circ}$ east of north. Table 
\ref{tab:cavities} summarizes the cavity properties and their derived energetics. The temperature, 
density, and pressure of the surrounding ICM is taken from Figure \ref{fig:deproj} at a radius 
corresponding to the centre of the cavities. Projected temperatures and deprojected densities are used 
in this analysis.

Cavity volumes are calculated using the geometric mean between oblate and prolate ellipsoids, 
$V=\frac{4}{3}\pi(ab)^{3/2}$. Upper (lower) limits on cavity volume are determined using the maximum 
(minimum) projected sizes, with volumes calculated in the same way as the mean volume. The total gas 
mass displaced by the cavities, assuming they are devoid of X-ray emitting material, is 
$M_{\rm disp}=2n_e\mu m_p V$. The electron density, $n_e$, is taken at a radius corresponding to the 
centre of the cavity. With volumes of $15\err{9}{6}\e{70}\cmcu$ and $16\err{7}{8}\e{70}\cmcu$ for the 
NE and SW cavities, respecticely, the displaced gas masses are $6\err{3}{2}\e{11}\Msun$ and $5\err{2}{3}\e{11}\Msun$. 
The $pV$ work required to inflate the cavities is $1.1\err{0.6}{0.4}\e{61}\erg$ for the NE cavity and 
$1.1\err{0.5}{0.5}\e{61}\erg$ for the SW cavity.

The cavity ages estimated using the buoyancy, sound crossing, and refill timescales are listed in Table 
\ref{tab:cavities}. The gravitational acceleration, $g$, used to calculate the buoyancy and refill timescales 
was determined using the MS0735 mass profile from Main et al. (in prep). In general the terminal velocity of 
bubbles is $\sim50\%$ of the sound speed \citep{rafferty06}. In MS0735, however, the large bubble volumes result 
in supersonic terminal velocities. Neglecting the bubble's expansion history in the buoyancy timescale therefore
 underestimates the true cavity age. We use the mean of the buoyancy, sound crossing, and refill timescales 
in order to estimate bubble age.

The ages of the NE and SW cavities are $1.5\e{8}\yr$ and $1.7\e{8}\yr$, respectively. The age of the 
surrounding shock front, $1.1\e{8}\yr$ (see Section \ref{sec:outer_shock}), which should be comparable 
to the true cavity age, is shorter than the mean rise time. We therefore expect to have slightly overestimated 
cavity ages, so the calculated power is likely underestimated. The power required to inflate these bubbles 
is $9\err{5}{4}\e{45}\ergps$ and $8\pm4\e{45}\ergps$ for the NE and SW cavities, respectively. The total 
enthalpy ($4pV$) of these cavities is $8.8\err{3.2}{2.4}\e{61}\erg$, and the total power is 
$1.7\err{0.6}{0.5}\e{46}\ergps$.

The total cavity power of this AGN outburst exceeds the X-ray bolometric luminosity within the 
cooling radius, $2.6\e{44}\ergps$, by more than a factor of $60$, easily compensating for radiative losses. 
At a projected distance of $\sim150\kpc$, the majority of the cavity volume is located outside of 
the $100\kpc$ cooling region. Unless adiabatic losses from cosmic ray streaming account for a 
significant fraction of energy dissipation, most of the cavity enthalpy will be carried outside 
of the cooling region. If this energy is deposited within $1\Mpc$, where the total gas mass is 
$7\e{13}\Msun$ (Main et al. in prep), the total cavity enthalpy heats the gas by $0.4\keV$ per particle.

\begin{figure}
\centering
\includegraphics[width=0.8\columnwidth]{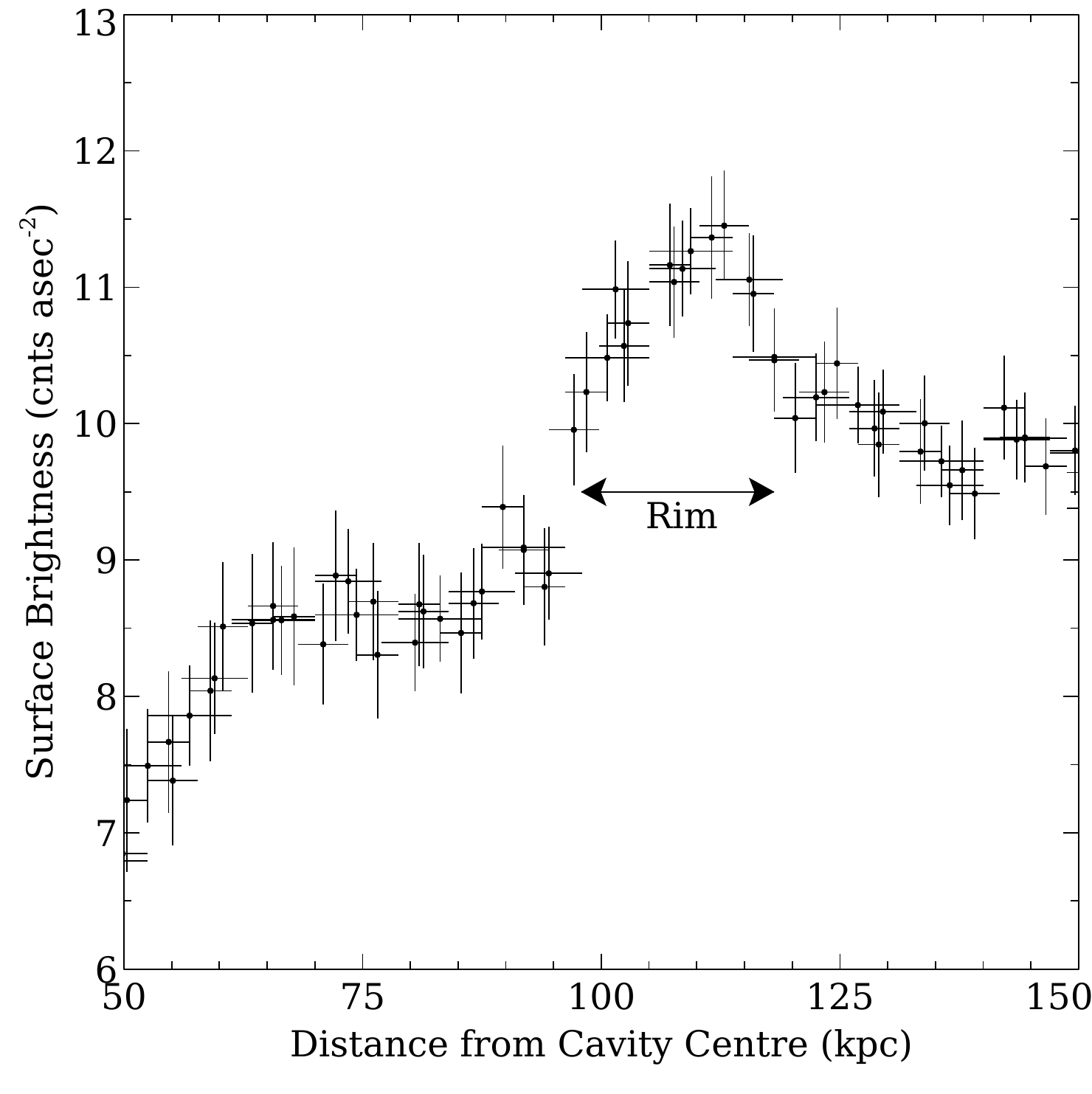}
\caption{Surface brightness cut centred on the NE cavity and extending to the west with an $80^{\circ}$ opening angle. Bright emission from rims surrounding the cavity is observed between $100$ and $120\kpc$. Profiles are shown for a variety of radial bin widths.}
\label{fig:rims}
\end{figure}

\subsection{Cavity Rims}
\label{sec:rims}

Bright rims of cool gas surround many cavity systems (e.g. \citealt{fabian00, nulsen02, blanton01, blanton11}), 
and are also observed in MS0735. The brightest rim appears along the western edge of the NE cavity. A 
surface brightness profile, centred on the cavity, was created in order to determine the width of the 
rim. This profile, shown in Figure \ref{fig:rims}, is created from sectors with an $80^{\circ}$ opening 
angle oriented to the west. Surface brightness is enhanced between $\sim100$ to $120\kpc$, giving a rim 
width of $20\kpc$.

A spectrum was extracted for the rim and fit with the absorbed thermal model discussed in Section \ref{sec:model}. 
For comparsion, a spectrum was extracted from a $20\kpc$ region just outside of the rim with the same opening 
angle. The rim contains cooler gas than the surrounding gas, with $kT=5.7\pm0.4\keV$ compared to the 
surrounding $6.8\err{0.6}{0.5}\keV$. The ratio between densities can be determined directly from the 
normalization of the thermal model, $K$, and the volume of each region,
\begin{equation}
\frac{n_{\rm rim}}{n_{\rm amb}} = \sqrt{\frac{K_{\rm rim}}{K_{\rm amb}}\frac{V_{\rm amb}}{V_{\rm rim}}}.
\end{equation}
We find that the rim is a factor of $1.18\pm0.03$ more dense than the surrounding gas. The pressure 
ratio between the rim and the ambient gas, $1.0\pm0.1$, implies that they are in pressure balance.

\begin{figure}
\centering
\includegraphics[angle=-90, width=0.97\columnwidth]{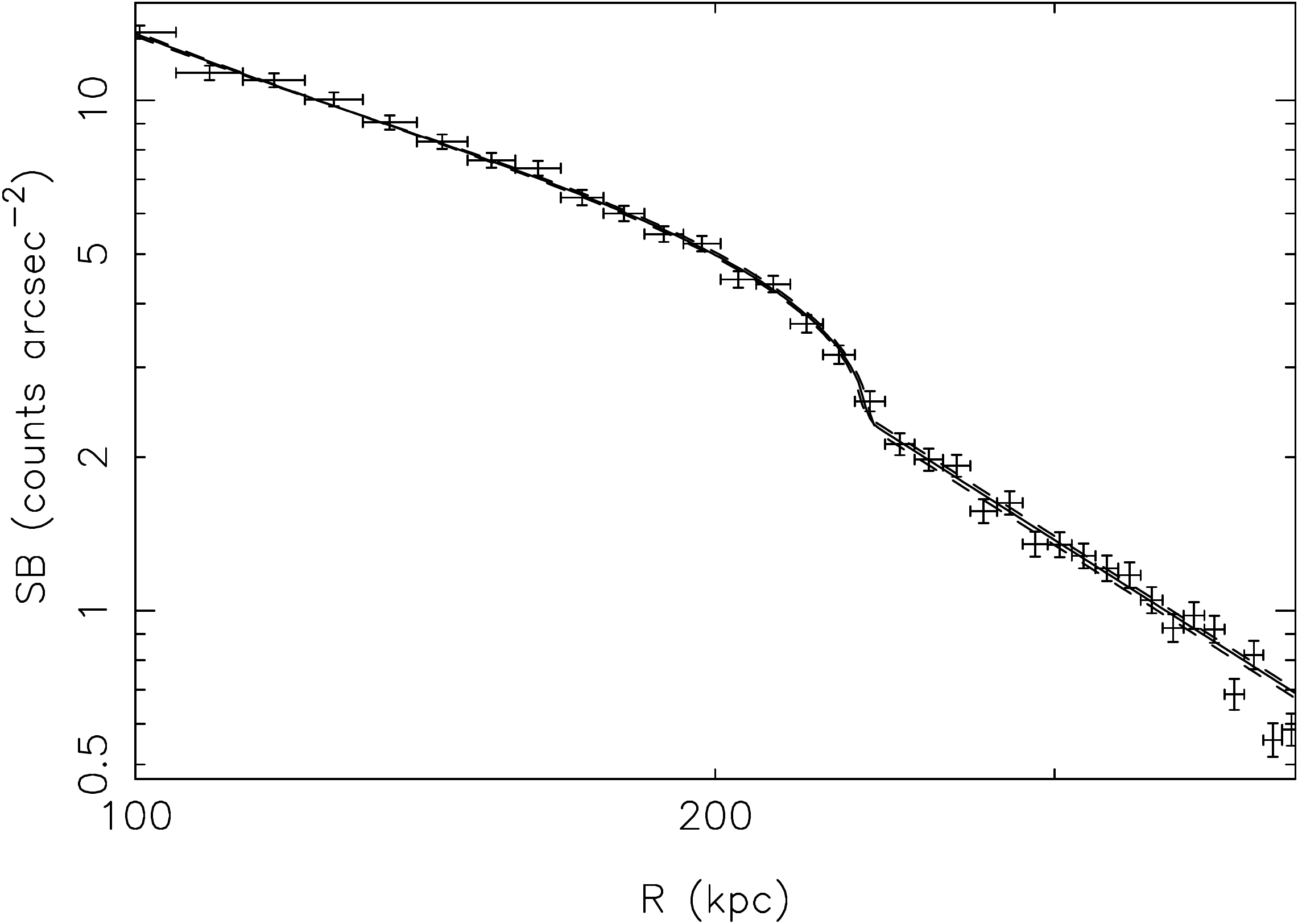}
\caption{
Projected surface brightness profile of the outer shock front compared to shock model predictions. The $0.5-7.0 \keV$ surface brightness is measured in sectors centred around the semi-minor axis of the shock front with opening angles of $30^{\circ}$ to the East and $30^{\circ}$ to the West. This surface brightness profile is best fit by a shock with Mach number $\mathcal{M}=1.26^{+0.04}_{-0.09}$ at a radius of $245\kpc$. The solid line shows the best fit from the shock model, which has been scaled to match the preshock surface brightness. The dashed lines show the fits from Mach numbers of $1.17$ and $1.30$.
}
\label{fig:outer_shock}
\end{figure}

\subsection{Shock}
\label{sec:outer_shock}

Radio jets launched by the AGN drive shock fronts into the ICM. MS0735 hosts a continuous, elliptical 
shock front that encompasses the outer pair of cavities \citep{nature05}. This shock front is modeled 
as a spherically symmetric point explosion in an intially isothermal atmosphere in hydrostatic equilibrium. 
The surface brightness profile computed from the model is scaled to fit the preshock profile in the 
specified energy band. The postshock conditions are then dictated by the Rankine-Hugoniot shock jump 
conditions for a given Mach number, $\mathcal{M}$. A detailed description of the analysis can be found 
in \citet{nulsenherc}.

The shock front is qualitatively fit by an ellipse with semi-major axis $a=320\kpc$, semi-minor 
axis $b=230\kpc$, and position angle of $7^{\circ}$ east of north. We approximate the elliptical 
shock front using spherical symmetry by considering only small opening angles around the minor axis 
of the shock ($30^{\circ}$, roughly to the east and west). 
We are unable to accurately model the shock along the major axis because of the depression in postshock surface brightness caused by the large cavities.
The $0.5-7.0\keV$ surface brightness 
profile extracted from these circular sectors is shown in Figure \ref{fig:outer_shock} and is best 
fit by Mach number $\mathcal{M}=1.26\err{0.04}{0.09}$ ($90\%$ confidence) at a radius of $245\kpc$. 
This value is marginally ($2\sigma$) below the $1.41\pm0.07$ measured by \citet{nature05}. The 
difference is likely related to the improved precision yielded by the deeper observation, which 
provides a more accurate measurement of the pre- and post-shock surface brightness profiles and 
a more accurate characterization of the depth of the surface brightness discontinuity.
The surface brightness of the preshock gas is best fit by the powerlaw $r^{-\beta}$ with $\beta=2.48\pm0.06$. 
The corresponding density profile, assuming constant temperature, is $r^{-\eta}$ with $\eta=1.74\pm0.04$.

\begin{figure}
\centering
\includegraphics[width=\columnwidth]{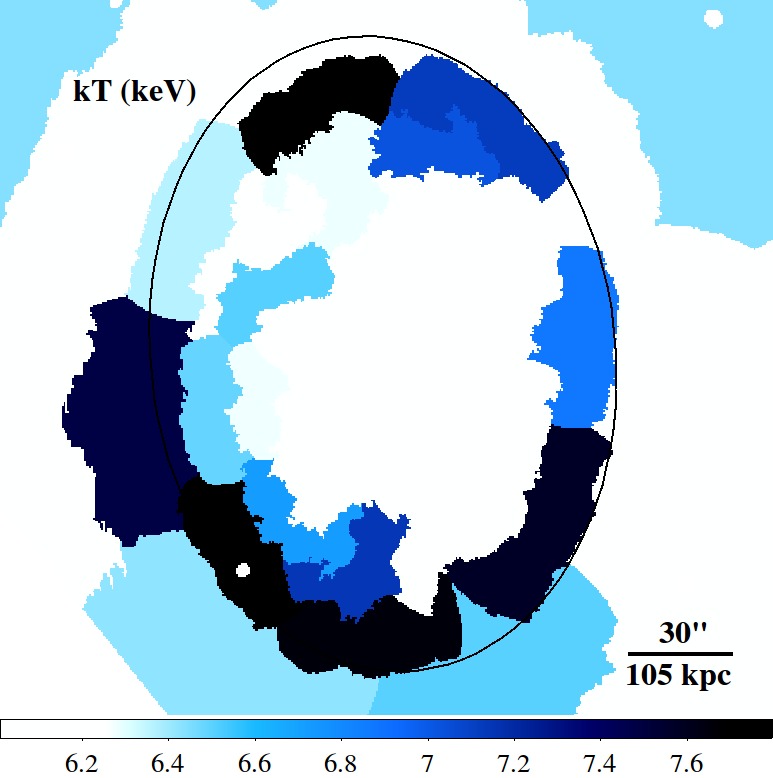}
\caption{Same temperature map as in Figure \ref{fig:maps}, but scaled to focus on the shock front. The postshock temperature is highest along the major axis of the shock front, where the Mach number is higher.}
\label{fig:postshock}
\end{figure}

From the Rankine-Hugoniot jump conditions, a Mach $1.26$ shock causes a $25\%$ increase in temperature. 
Projection effects obscure the jump, so that the expected temperature jump is only $\sim10\%$. 
From Figures \ref{fig:profiles} and \ref{fig:deproj} the observed temperature jump, $1.14\pm0.07$, 
is within $1\sigma$ of the expected jump. A map of the postshock temperature is provided in Figure 
\ref{fig:postshock}, with the location of the shock front shown as the black ellipse. 
The postshock temperature is clearly highest along the major axis of the shock, reaching $\sim8\keV$ to 
both the North and South but only $\sim7\keV$ to the East and West. This is expected of an elliptical 
shock front, since the major axis must propagate faster and therefore has a higher Mach number. Due to 
the large uncertainties in temperature, the jumps along the major and minor axes are not significantly 
different. However, Figure \ref{fig:postshock} shows a clear azimuthal trend in temperature, which is 
indicative of an enhanced temperature jump along the major axis of the shock front.

The shock energy is determined from the point explosion model and depends on preshock temperature 
and density. The preshock properties are taken from the elliptical profiles in Figure \ref{fig:profiles}, 
which trace the shock front well. The preshock temperature is specified as $6.5\keV$ with a density 
of $1.85\e{-3}\cm^{-3}$ at $250\kpc$, resulting in a shock energy of $4\e{61}\erg$. Using the deprojected 
profiles in Figure \ref{fig:deproj} to determine the preshock conditions do not change the result, 
which is likely only accurate within a factor of $\sim2$. 

The main shortcomings of the shock model are in the assumptions of spherical symmerty and a 
point explosion. Spherical symmetry is a poor description of the elliptical shock front, so 
the shocked volume is underestimated. The AGN likely drives the shock through a continuous 
injection of energy, which requires more energy than a point explosion in order to generate 
the same shock strength. The energy measured here, $4\e{61}\erg$, is therefore an underestimate 
of the true shock energy. An improved shock model that does not assume spherical symmetry or 
a point explosion is required in order to improve the analysis of this weak shock.

The age of the shock determined by the model, $1.1\e{8}\yr$, is comparable to the buoyancy 
and sound-crossing times of the cavities. Since they originate from the same AGN outburst, 
these ages are likely close to the true value. The shock power, $E_{\rm shock}/t_{\rm shock} \approx 1.1\e{46}\ergps$, 
is slightly smaller than the total cavity power of $1.7\e{46}\ergps$. Exceeding the X-ray luminosity 
within the cooling region by a factor of $40$, the shock front also possesses ample power to offset 
radiative losses. Combined, the cavities and shock front are able to heat the ICM by $0.6 \keV$ 
per particle within $1\Mpc$, which provides a significant fraction of the $1-3\keV$ per particle 
required to preheat the cluster gas \citep{wu00}.

\begin{figure*}
\begin{minipage}{0.95\textwidth}
\centering
\includegraphics[width=0.45\columnwidth]{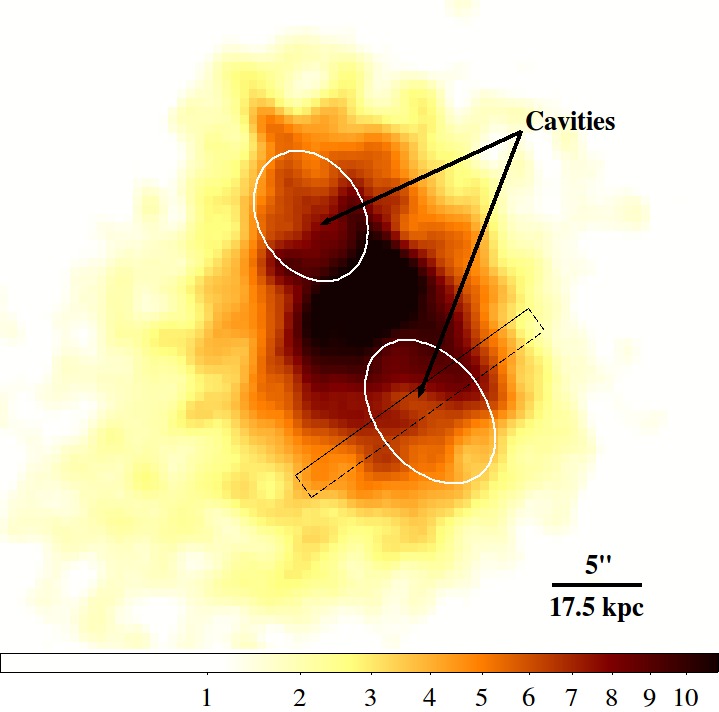}
\includegraphics[width=0.4\columnwidth]{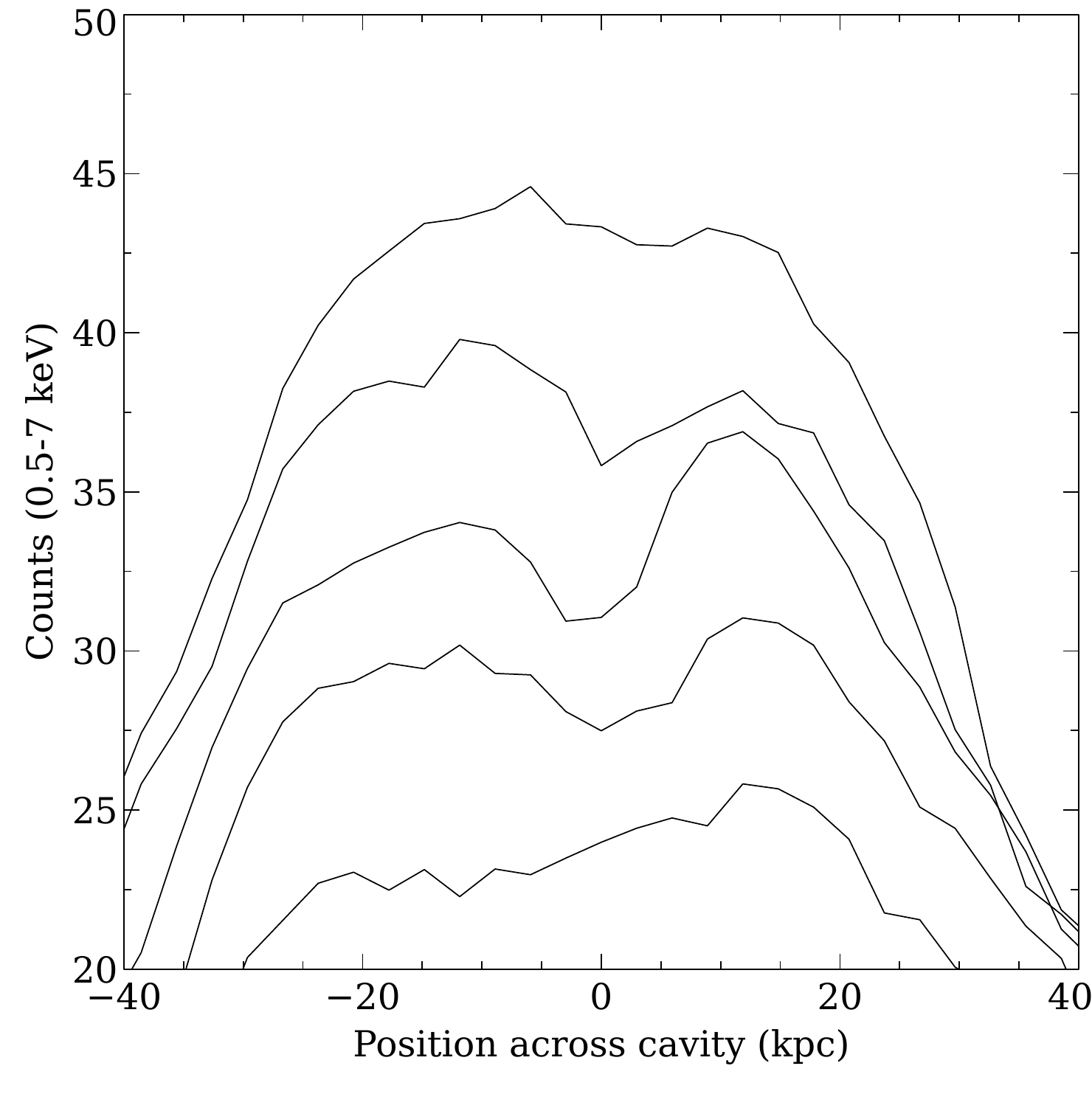}
\end{minipage}
\caption{
\textit{Left}: Soft band ($0.3-1.0\keV$) image in units of counts pixel$^{-1}$, Gaussian-smoothed with a $1.5''$ kernel radius. Two regions of low surface brightness are visible in this image, which we interpret as a pair of cavities originating from a recent AGN outburst.
\textit{Right}: Counts from the $0.5-7\keV$ band for a series of linear projections lying across the SW inner cavity. The projections are taken at a series of radii from the cluster centre, lying roughly perpendicular to the radial direction. The region associated with the middle profile is shown in the left panel as the rectangle. The position across the cavity starts from the NW and runs toward the SE.
}
\label{fig:inner_cavities}
\end{figure*}

\section{Rejuvenated Outburst}
\label{sec:smalloutburst}

The deep X-ray image reveals two smaller cavities or channels located along the radio jets in the inner $20\kpc$ of the
BCG. Due to their small size relative to the larger cavities and their proximity to the cluster core, 
the cavities are difficult to disentangle from the gas presumably displaced by the large cavity system. 
The presence of a second pair of cavities is indicative of a rejuvenated AGN outburst. The inner edge 
of these cavities are traced by cool rims of displaced gas that are evident in the soft band ($0.3-1.0 \keV$) 
image, shown in Figure \ref{fig:inner_cavities} (left). This image was prepared in the same way as 
Figure \ref{fig:ms07}, but with energies binned between $0.3$ and $1.0\keV$.

The right panel of Figure \ref{fig:inner_cavities} shows five projections taken across the SW 
inner cavity. The black rectangle in the left panel of Figure \ref{fig:inner_cavities} shows 
the projection corresponding to the middle profile. Counts in the $0.5-7\keV$ energy band were 
averaged over the $5\kpc$ width of the projection for each pixel along its length. Several of 
the profiles show a decrease in brightness, the most significant of which occurs $\sim22\kpc$ 
from the cluster centre and reveals a $10-20\%$ deficit in counts relative to the surrounding 
bright rims. The decrease in counts toward the NE cavity is much smaller than is seen in the 
SW cavity, merely flattening instead of producing a clear deficit. This difference likely results 
from the projection into our line of sight of emission from dense gas, nonuniformly distributed in the cavity rims.

The projected sizes of the two inner cavities are determined in the same manner as the outer 
cavities (see Section \ref{sec:cavities}). The white ellipses in Figure \ref{fig:inner_cavities} 
(left) show the cavity sizes adopted here, which use the middle of the rims to determine the mean 
cavity volume. This approach gives a projected size of $a=13.3\err{1.1}{2.0}\kpc$, 
$b=10.1\err{2.5}{2.1}\kpc$ for the inner NE cavity and $a=15.5\err{2.3}{2.0}\kpc$, $b=10.5\err{3.2}{3.3}\kpc$ 
for the inner SW cavity. Volumes for these cavities are calculated using the geometric mean between oblate 
and prolate ellipsoids, or $V=\frac{4\pi}{3}(ab)^{3/2}$. At projected distances of $19.3$ and $25.0 \kpc$ 
from the cluster centre, these cavities displace $8.2\e{9}\Msun$ and $7.7\e{9}\Msun$ of gas in the ICM, respectively.

The mechanical energy required to inflate these cavities, $pV$, is $9\err{5}{4}\e{58}\erg$ for the inner 
NE cavity and $9\err{7}{5}\e{58}\erg$ for the inner SW cavity. The mean age of the bubbles, $4.8\e{7} \yr$ 
and $4.2\e{7} \yr$, give cavity powers of $2.8\err{1.6}{1.3}\e{44}\ergps$ and $2.4\err{2.0}{1.3}\e{44}\ergps$,
respectively. The total mechanical energy ($4pV$) and power for this outburst are therefore 
$7.2\err{3.4}{2.6}\e{59}\erg$ and $5.2\err{2.6}{1.8}\e{44}\ergps$. This cavity power exceeds 
radiative losses within the cooling region ($2.6\e{44}\ergps$) by a factor of $2$ and is $30$ times 
smaller than the outer cavity power, implying that AGN power varies significantly over time. While 
feeble compared to the outer cavities, the inner cavities are comparable in power to the bubbles in Perseus.

Despite the ample energy in the outer outbursts, a rejuvenated AGN outburst is required in order to 
continue to offset radiative losses (see e.g. Perseus: \citealt{fabian06}, M87: \citealt{forman07}, 
Hydra A: \citealt{nulsenhydra}). The outburst interval in MS0735, given by the difference in mean bubble 
ages, is $1.1\e{8}\yr$. This value ranges between a minimum of $6\e{7}\yr$ when the buoyancy time is used 
to a maximum of $1.7\e{8}\yr$ with the refill timescale. Each of these values is shorter than the central 
cooling time, $6.4\err{0.9}{0.8}\e8\yr$. The AGN outbursts therefore occur on short enough timescales to 
prevent the majority of the ICM from cooling. Repeated heating of the central gas supplies enough energy 
to prevent cooling, and may explain the lack of star formation in the system \citep{mcnamara09}.

\subsection{Spectrally Hard Features: An Inner Shock?}
\label{sec:inner_shock}

\begin{figure}
\centering
\includegraphics[width=0.99\columnwidth]{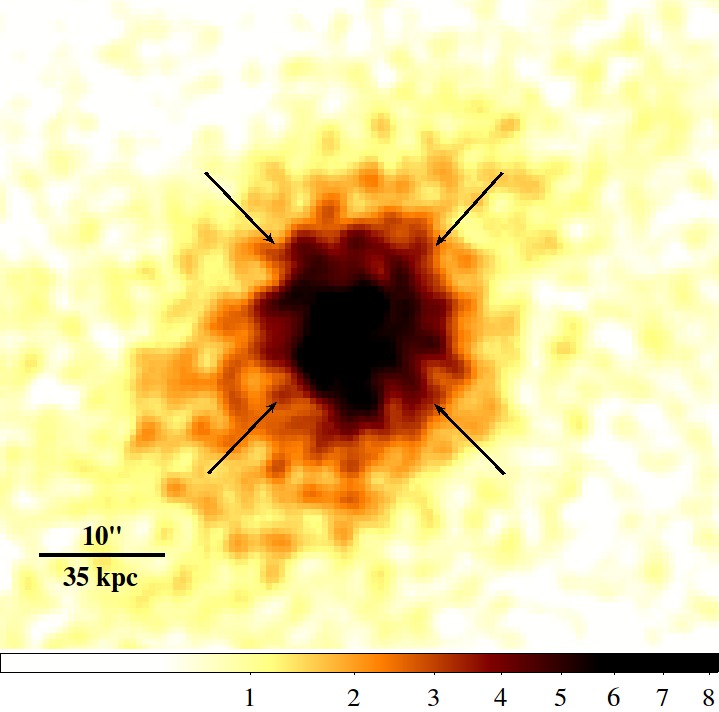}
\caption{
Hard band ($3-7\keV$) image in units of counts pixel$^-1$, smoothed with a two-dimensional Gaussian with a $1.5''$ kernel radius. The arrows indicate a circular feature that is reminiscent of the weak shock in M87.
}
\label{fig:hardband}
\end{figure}

Weak shocks are typically produced alongside cavities in an AGN outburst. The gas warmed by shock heating 
can be traced using a hard band image \citep{forman07}, which excludes the portion of the X-ray spectrum that is 
only weakly dependent on temperature (see Figure 1 of \citealt{arnaud05}). We produce a hard band image 
by binning cluster emission between $3$ and $7\keV$. 
The hard emission, shown in Figure \ref{fig:hardband}, is fairly circular, with 
extended emission lying perpendicular to the jet axis. This morphology is quite different than the 
cooler gas traced by the soft band (Figure \ref{fig:inner_cavities}), which shows emission extending along the jet axis.

The arrows in Figure \ref{fig:hardband} point to a circular feature that is reminiscent of the 
shock front in M87 \citep{forman07}. A jump in surface brightness is observed at this location, 
but it is marginal and depends strongly on radial binning. We therefore cannot conclusively 
determine if this feature corresponds to a shock front. If it is a shock front, then, approximating 
the density jump from the surface brightness jump, it would have a Mach number of approximately 
$1.1$. With a radius of $30\kpc$ the energy of the shock front is $\sim3\e{59}\erg$, which is a 
factor of $2$ smaller than the total mechanical energy required to inflate the associated bubbles. 
The age of this shock is $2.6\e{7}\yr$ for a preshock temperature of $4\keV$, which translates to 
a power output of $\sim4\e{44}\ergps$. This shock power is comparable to the total cavity power 
for the rejuvenated outburst as well as the X-ray bolometric luminosity within the cooling region.

\section{AGN Outburst Time Dependence}

MS0735 is one among several clusters and groups with two or more pairs of radio cavities 
embedded in their hot atmospheres. These systems are particularly interesting because they 
permit a direct indication of AGN power variability over time and the time interval between 
outbursts, which should be shorter than the cooling time if AGN are regulating cooling 
\citep{dunn06,rafferty08,birzan12}. Those known to us are listed in Table \ref{tab:sample}. 
We compare the time interval between outbursts in these systems to central cooling time and 
cavity heating time. We note that this archival sample is heavily biased by selection, as 
only large bubbles are detectable and multiple outbursts can only be detected for a select 
range of outburst intervals. There is likely a range in cavity sizes and ages that we are 
currently unable to detect \citep{review07}.

Outburst interval is determined from the difference in ages between inner and outer cavities, 
with buoyancy times used to determine bubble ages. Outburst intervals are also determined 
from the period of weak shocks and sound waves. A factor of $2$ uncertainty is attributed to 
all outburst intervals. For bubbles this roughly translates to calculating bubble ages from 
the sound crossing time and refill time and using those estimates as lower and upper bounds, 
respectively. The outburst intervals presented here, determined from cavity ages, range from 
$6\e6\yr$ in M87 to $1.8\e8\yr$ in Hydra A. The mean outburst interval of these points is 
$6\err{2}{1}\e{7}\yr$, while sound waves occur on much shorter timescales, with a mean period 
of $0.7\err{0.5}{0.2}\e{7}\yr$.

\begin{table}
\caption{Sample of systems with multiple AGN outbursts.}
\begin{minipage}{\columnwidth}
\begin{center}
\begin{tabular}{l c c c}
\hline
Cluster & Outburst Type\footnote{Type of outburst used to calculate outburst interval. 1: Cavities, 2: Shocks or sound waves.} & Reference \\
\hline
Perseus         & 1, 2 & \citet{fabian06} \\
M87             & 1, 2 & \citet{forman07} \\
MS 0735.6+7421  & 1    & This work \\
Hydra A         & 1    & \citet{wise07} \\
HCG 62          & 1    & \citet{rafferty13} \\
A2052           & 1    & \citet{blanton09, blanton11} \\
A2199           & 1    & \citet{nulsen13} \\
NGC 5813        & 1    & \citet{randall11} \\
NGC 5846        & 1    & \citet{machacek11} \\
A3581           & 1    & \citet{canning13} \\
Centaurus       & 2    & \citet{sanders08} \\
\hline
\end{tabular}
\end{center}
\end{minipage}
\label{tab:sample}
\end{table}

\begin{figure}
\centering
\includegraphics[width=\columnwidth]{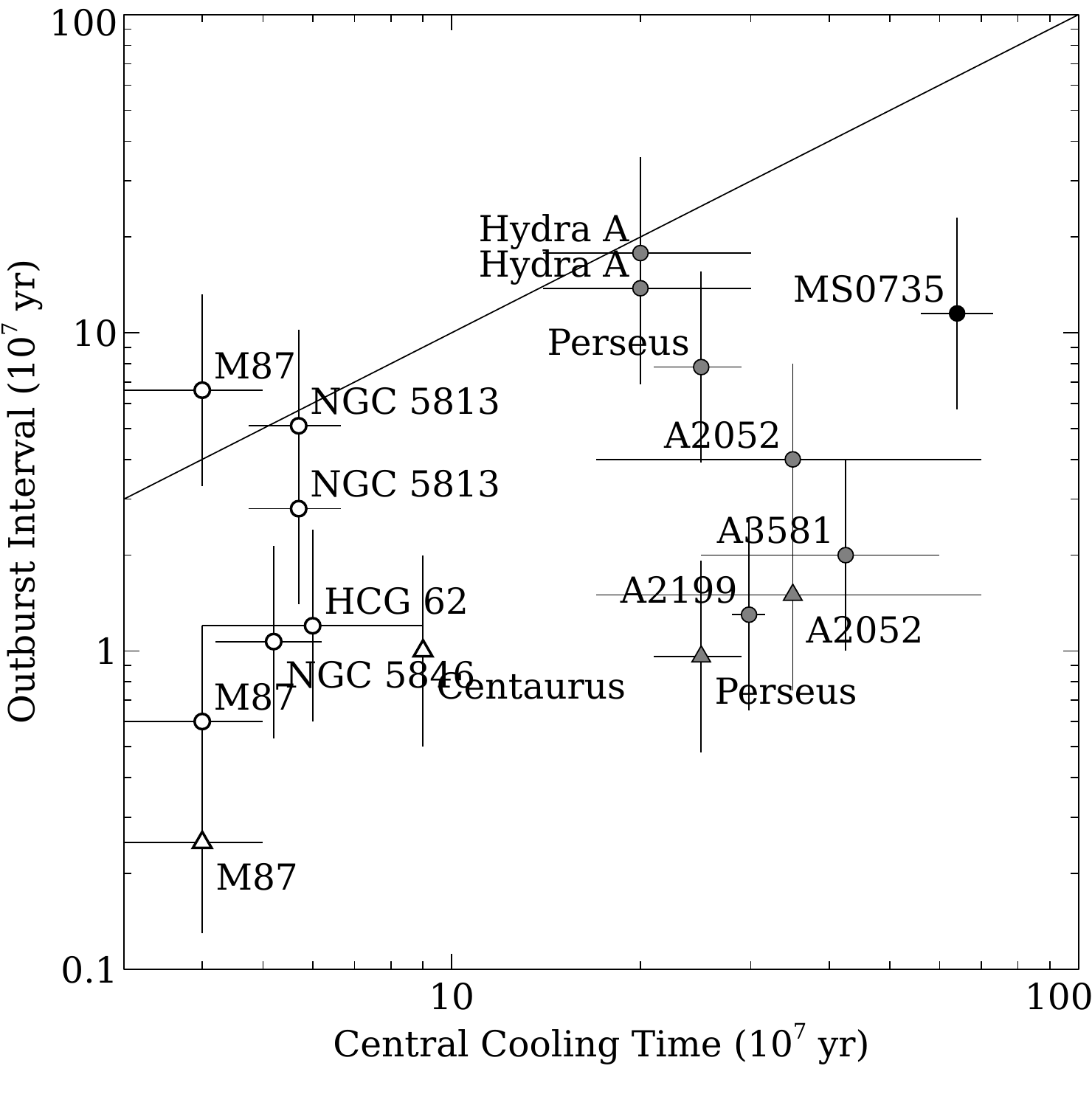}
\caption{AGN outburst interval plotted against central cooling time. Circles represent outburst intervals obtained from cavity ages while triangles are the estimates from sound waves or repeated weak shocks. The points are colour-coded based on the resolution, where white points are cooling times from the inner $1 \kpc$, grey are from the inner $5 \kpc$, and black are from the inner $10 \kpc$. In each system the duty cycle is shorter than the central cooling time. The line of equality between duty cycle and central cooling time is shown.
}
\label{fig:dutyvscool}
\end{figure}

Central cooling times are obtained from the \citet{rafferty08} and \citet{cavagnolo09} samples, 
and are plotted against outburst interval in Figure \ref{fig:dutyvscool}. Outburst intervals 
determined from bubble ages are plotted as circles, while those determined from weak shocks or 
sound waves are plotted as triangles. The points are colour-coded based on the central resolution 
element. Nearby systems, where the achievable resolution is $<1\kpc$, are shown in white. Grey points 
correspond to a resolution of $1-5\kpc$ and black points are for systems with a resolution that 
exceeds $5\kpc$. The clear separation between these points can be attributed to resolution \citep{peres98}. 
A consistent physical scale would shift the grey and black points to the left, but they tend to be 
hotter systems so would still have a longer central cooling time than most of the white points. 
Multiple points are shown for Hydra A, M87, and NGC 5813 because they host bubbles from 3 AGN 
outbursts, corresponding to 2 outburst intervals.

After accounting for resolution, Figure \ref{fig:dutyvscool} shows that outburst interval is 
shorter than or consistent with the mean central cooling time for each system in our sample. 
AGN outbursts therefore occur on short enough timescales to continually suppress cooling flows 
(see \citealt{rafferty08}). The mean cooling time for this sample of groups and clusters is 
$19\err{4}{2}\e{7}\yr$, which is a factor of $3$ larger than the mean outburst interval. The 
white points in Figure \ref{fig:dutyvscool} have a much shorter mean cooling time, $5.0\err{0.7}{0.6}\e{7}\yr$, 
than the points in grey, $30\err{8}{5}\e{7}\yr$. These points also have shorter outburst intervals, 
with a mean outburst interval of $2.9\err{1.5}{0.7}\e{7}\yr$ compared to $8\err{4}{2}\e{7}\yr$. 
Therefore the points with the smallest resolution element, which have shorter central cooling times, 
also have shorter outburst intervals. The period of sound waves in M87 \citep{forman07}, 
Perseus \citep{fabian06}, Centaurus \citep{sanders08}, and A2052 \citep{blanton09} are each an 
order of magnitude smaller than their central cooling time, implying that weak shocks repeat 
on timescales short enough to offset cooling in cluster centres.

\begin{figure}
\centering
\includegraphics[width=\columnwidth]{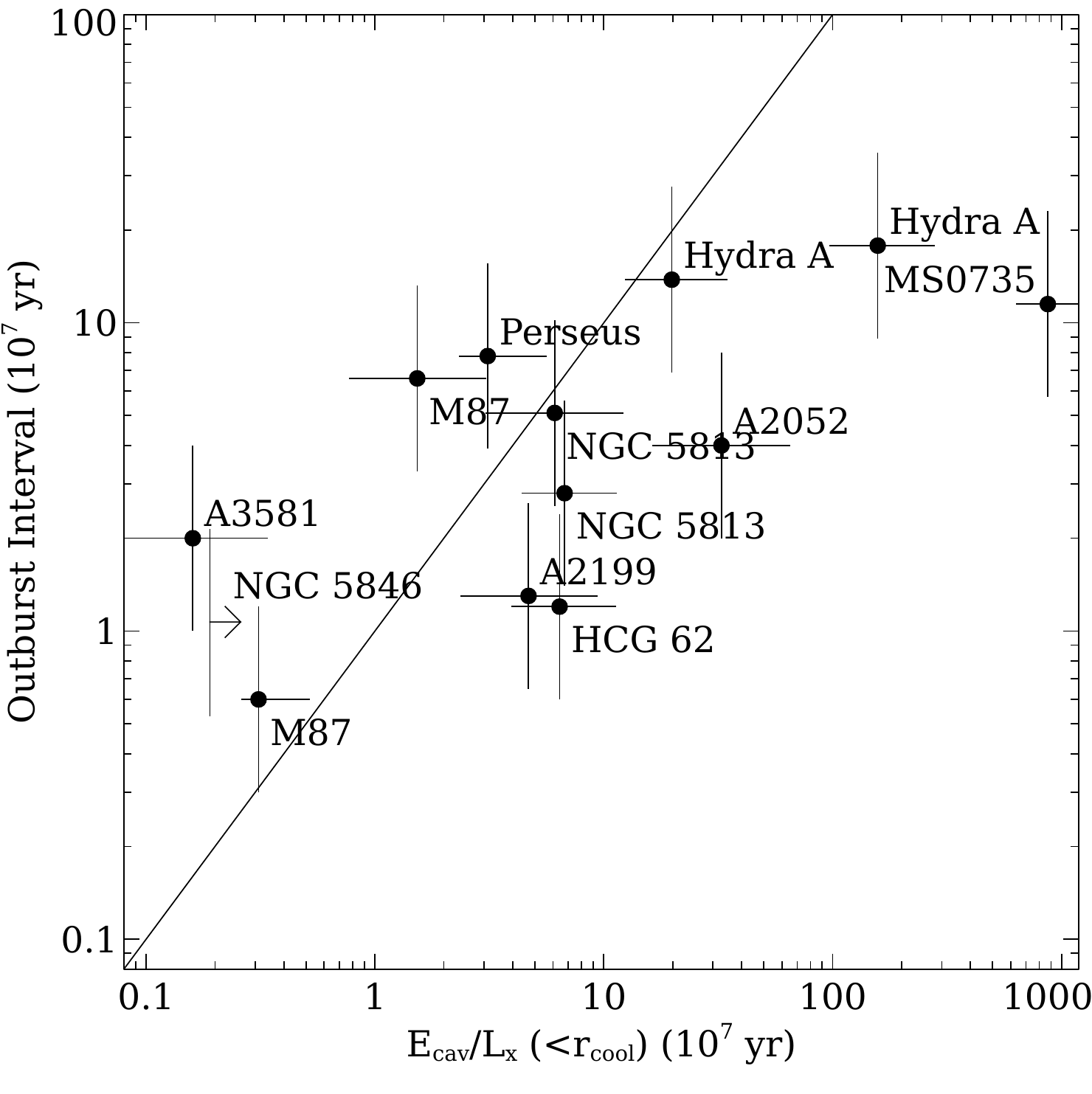}
\caption{AGN outburst interval plotted against heating time -- the amount of time that the outer cavity is able to compensate for radiative losses. Only outburst intervals determined from cavities are included in this plot. The solid line is the line of equality between the two timescales.
}
\label{fig:dutyvspV}
\end{figure}

We define cavity heating time as the amount of time that cavity enthalpy is able to offset 
radiative losses, $t_{\rm heating}=4pV/L_{\rm X}(< r_{\rm cool})$. A large outburst offsets 
cooling for longer times, which could translate to a longer outburst interval. Cavity heating 
time is compared to outburst interval in Figure \ref{fig:dutyvspV}. For the factor of $30$ range 
in outburst intervals, cavity heating time spans $4$ orders of magnitude. We find that outburst 
interval is shorter than cavity heating time for $8$ of the $13$ points in our sample. We note 
that $t_{\rm heating}$ is likely underestimated in NGC 5846 by a factor of a few, as 
\citet{machacek11} report the X-ray luminosity within $200 \kpc$ while the cooling radius 
is approximately $25\kpc$ \citep{cavagnolo09}. 

The majority of systems in Figure \ref{fig:dutyvspV} (8 of 13) have an outburst interval 
shorter than their cavity heating time, indicating that heating is continually able to offset 
radiative cooling over the age of the cluster. Furthermore, the presence of weak shocks and 
sound waves imply that cavity enthalpy is only a lower limit on the total energy of an AGN outburst, 
so an AGN outburst will be able to heat the system for longer than the heating time shown in 
Figure \ref{fig:dutyvspV}. The very small range in outburst intervals relative to the $4$ order 
of magnitude range in heating times suggests that any relationship between these timescales is 
weak. A much larger sample is required in order to draw any conclusions. The current sample is 
heavily biased toward nearby systems with large outbursts, and only a small range in outburst interval is detectable.

\section{Summary}

We have presented an analysis of a deep \textit{Chandra} observation of MS0735. We find that 
the mean power required to inflate the large cavity system is $1.7\e{46}\ergps$, which is the 
most energetic outburst known. These cavities are encompassed by a continuous, elliptical 
shock front. We fit a simple hydrodynamical model to the surface brightness profile and obtain 
a Mach number of $1.26\err{0.04}{0.09}$. The mean power required to drive this shock front, 
$1.1\e{46}\ergps$, is comparable to the power partitioned to the cavities. A clear temperature 
jump is associated with the shock front, and appears to be stronger
along its major axis, where the Mach number must be higher. The power of the AGN outburst is 
more than enough to offset the modest cooling luminosity of $2.6\e{44}\ergps$.

We report the detection of a pair of bubbles corresponding to a more recent AGN outburst. 
These bubbles are located at a projected distance of $20-25\kpc$ from the cluster centre, 
where the outer cavities are no longer able to heat the ICM. The mean power required to 
inflate these cavities is $5.2\e{44}\ergps$. A circular feature is observed in the hard band 
surface brightness, and resembles the weak shock front in M87. The associated surface brightness 
jumps are marginal, so we cannot conclude with certainty if this is, in fact, a shock front. 
If it is a shock front, it has a Mach number of approximately $1.1$ and a mean power of $4\e{44}\ergps$.

We detect multiphase gas, with temperatures of $3.4$ and $0.65\keV$, within $14\kpc$ of the 
cluster centre. Evidently, little of this gas condenses out of the intracluster medium. The 
mean power of the inner cavities is a factor of $40$ larger than the combined X-ray luminosities 
of the two gas phases, and is therefore powerful enough to suppress cooling. Alternatively, the 
condensation of gas out of the ICM can be prevented by removing the supply of low entropy gas from 
the centre of the cluster. Cool gas is observed preferentially along the direction of the radio 
jet, indicating that gas has been entrained by the jet and is being dragged to high altitudes.

With the addition of MS0735, we prepare a sample of $10$ clusters and groups with multiple generations 
of cavities. We also include three systems with detected sound waves, two of which overlap with 
the cavity sample. We find that the outburst interval is shorter than central cooling time in every 
system. We also compare outburst interval to the heating time, defined as $4pV/L_{\rm X} (< r_{\rm cool})$. 
We find that the majority of systems in this analysis have a heating time longer than their outburst 
interval, implying that AGN cycling is able to continually suppress cooling.

\section*{Acknowledgements}
We thank the anonymous referee for helpful comments that improved the paper. ANV thanks Julie 
Hlavacek-Larrondo for helpful comments. ANV, BRM, HRR, and RAM acknowledge support 
from the Natural Sciences and Engineering Research Council of Canada. BRM acknowledges funding from 
the Chandra X-ray Observatory Cycle 10 Large Project proposal for MS0735.

\bibliographystyle{mn2e}
\bibliography{ms07}

\end{document}

%% file: defn.tex
\newcommand{\Mpc}{\rm\; Mpc}
\newcommand{\kpc}{\rm\; kpc}

\newcommand{\km}{\rm\; km}

\newcommand{\cm}{\rm\; cm}

\newcommand{\cmsq}{\hbox{$\cm^2\,$}}
\newcommand{\cmcu}{\hbox{$\cm^3\,$}}

\newcommand{\yr}{\rm\; yr}

\newcommand{\Myr}{\rm\; Myr}
\newcommand{\s}{\rm\; s}

\newcommand{\ks}{\rm\; ks}

\newcommand{\MHz}{\rm\; MHz}

\newcommand{\Msun}{\hbox{$\rm\thinspace M_{\odot}$}}

\newcommand{\Msunpyr}{\hbox{$\Msun\yr^{-1}\,$}}

\newcommand{\keV}{\rm\; keV}

\newcommand{\erg}{\rm\; erg}

\newcommand{\ergps}{\hbox{$\erg\s^{-1}\,$}}

\newcommand{\kmps}{\hbox{$\km\s^{-1}\,$}}

\newcommand{\kmpspMpc}{\hbox{$\kmps\Mpc^{-1}\,$}}

\newcommand{\Zsun}{\hbox{$\thinspace \mathrm{Z}_{\odot}$}}

\newcommand{\psqcm}{\hbox{$\cm^{-2}\,$}}

\newcommand{\pcmcu}{\hbox{$\cm^{-3}\,$}}

